\documentclass[%
 amsmath,amssymb,
 aps,
pra,
]{revtex4-2}

\usepackage{graphicx}
\usepackage{dcolumn}
\usepackage{epsfig}
\usepackage{graphics,color}
\usepackage{amssymb}
\usepackage{amsmath}
\usepackage{latexsym}
\usepackage{hyperref}
\hypersetup{colorlinks=true, citecolor=blue, urlcolor=blue, linkcolor=red}
\usepackage{amsbsy}
\usepackage{bm}
\usepackage{url}

\begin{document}
\title{Bubble dynamics in the Polyakov quark-meson model}
\author{Junrong Wang}
\author{Jinshuang Jin}
\author{Hong Mao}
\email {mao@hznu.edu.cn (corresponding author)}
 \affiliation{1. School of Physics, Hangzhou Normal University, Hangzhou 311121, China}


\begin{abstract}
In the framework of the Polyakov quark-meson model with two flavors, the bubble dynamics of a first-order phase transition in the region of high density and low temperature are investigated by using the homogeneous thermal nucleation theory. In mean-field approximation, after obtaining the effective potential with inclusion of the fermionic vacuum term, we build a geometric method to search two existing minima, which can be actually connected by a bounce interpolated between a local minimum to an adjacent global one. For both weak and strong first-order hadron quark phase transitions, as fixing the chemical potentials at $\mu=306 \mathrm{MeV}$ and $\mu=310 \mathrm{MeV}$, the bubble profiles, the surface tension, the typical radius of the bounce and the saddle point action as a function of temperature are numerically calculated in the presence of a nucleation bubble. It is found that the surface tension remains a very small value even when the density is high. It is also noticed that the deconfinement phase transition does not change chiral phase transition dramatically for light quarks and phase boundaries for hadron and quark matter should be resized properly according to the saddle-point action evaluated on the bounce solution.

\end{abstract}


\maketitle

\section{Introduction}

At extremely high temperatures and densities, a strong interaction theory, quantum chromodynamics (QCD), favours an idea that there is a phase conversion from a hadronic matter at low energy into a deconfined and chiral-symmetry restored quark matter denoted as quark-gluon plasma (QGP). In experiment, how to explore and illuminate the fundamental properties of QGP is a current topic concerned with the physics of the heavy-ion collisions (HICs) in laboratories \cite{Yagi:2005yb,Fukushima:2010bq,Braun-Munzinger:2015hba} as well as to the structures of compact stars in astrophysics \cite{Schaffner-Bielich:2020psc,Schmitt:2010pn,Baym:2017whm}. On the theoretical side, although QCD is applicable for determining the physical properties of strong-interaction matter at very high energy, to understand the nature of the hadron-quark phase transition remains as a challenge problem in actual calculations, in particular when finite chemical potentials are related. Since an \textit{ab initio} approach, Monte Carlo calculations on the lattice, is still severely hampered by the notorious fermion sign problem and can not be applied to study the QCD phase transition at high density \cite{Aarts:2023vsf}. Therefore, effective models of QCD provides an alternative therapy to capture the basic characteristics of QCD, such as two salient features intimately connected with the nonperturbative properties of QCD, chiral symmetry and the confinement. Among these effective models, some are the Nambu-Jona-Lasinio (NJL) model \cite{Nambu:1961tp,Nambu:1961fr,Bentz:2001vc,Hutauruk:2023mjj,Gifari:2024ssz,Hatsuda:1994pi,Buballa:2003qv}, the linear sigma model (LSM) \cite{Gell-Mann:1960mvl}, and their modern extensions included with effects of the deconfinement, the Polyakov Nambu-Jona-Lasinio model (PNJL) \cite{Fukushima:2003fw,Fukushima:2017csk,Costa:2010zw}, and the Polyakov Quark Meson Model (PQM) \cite{Schaefer:2007pw,Mao:2009aq,Schaefer:2009ui} for quark matter. While others are the relativistic mean field (RMF) model and its extrapolations incorporated with chiral symmetry, the chiral nucleon meson model  \cite{Floerchinger:2012xd,Drews:2013hha,Fraga:2018cvr,Schmitt:2020tac}, the parity doublet model \cite{Detar:1988kn,Gallas:2009qp,Zschiesche:2006zj,Fukushima:2013rx,Benic:2015pia,Koch:2023oez} and the relativistic mean field including both the chiral potential and the confinement effect \cite{Chanfray:2000ha,Chanfray:2005sa,Somasundaram:2021hna} for nuclear matter.     

After the observation of gravitational waves by LIGO Collaboration \cite{LIGOScientific:2016aoc}, we enter a new era of gravitational wave astronomy to directly study cosmological first-order phase transition predicted by the standard model \cite{Cutting:2018tjt,Cutting:2020nla,Athron:2023xlk}. The topic has gathered a renewed interest due to the fact that the present research in the cosmological first-order phase transition can be understood by detecting a stochastic gravitational wave background in the present or future gravitational wave experiments \cite{Caprini:2019egz,Hindmarsh:2020hop,Croon:2023zay}. Since the gravitational wave can reveal more information about the cosmological phase transition, once the signal of the stochastic gravitational wave spectrum is to be confirmed, it would largely increase our capability to probe the earliest Universe. At the same time, with the direct detection of gravitational waves from the binary neutron star merger event GW170817 \cite{LIGOScientific:2017vwq,LIGOScientific:2018hze}, astrophysics has also arrived at a new multimessenger age. Integrated with other observational astronomy, such as the radius, the mass and the tidal deformability of compact stars, there exists so far the most stringent constraint on the equation of states (EOS) of dense QCD Matter, which play a key role in the dynamical evolution and structure formation of compact stars \cite{Llanes-Estrada:2019wmz,Xia:2019pnq,Baiotti:2019sew,Blacker:2020nlq,Cao:2018tzm}. So that any efforts to study strong interaction as well as the possible first-order QCD phase transition are continuously deserved to be made in the field of astrophysics and cosmology.      

For QCD phase transition, most models in description of strong interactions predict a first-order phase transition from a chiral symmetry broken hadron phase to a chiral symmetric quark phase at high chemical potential and low temperature. Whereas, it is a smooth crossover conversion at low chemical potential but high temperature. Along the first-order coexistence line, there could exist a critical endpoint (CEP) where the first-order phase transition is to be of the second-order. In laboratories, to explore and locate the CEP through ultra-relativistic heavy-ion collision are the ultimate targets of the Beam Energy Scan (BES) programs at Relativistic Heavy-Ion Collider (RHIC) and other near future experimental facilities, the empirical identification of the CEP will be a milestone of studying QCD phase diagram and strong interactions\cite{Luo:2017faz,Pandav:2022xxx}. On the theoretical side, the presentation of the CEP guarantees the existence of the first-order phase transition, and vice versa. Since much attentions have been drawn to the crossover phase conversion of QCD over past decades, in the present work, most of our studies are focused on the topic of a first-order hadron quark phase conversion in QCD phase diagram.

For a typically first-order phase transition, a system will goes from a metastable high energy state (false vacuum) to a relatively stable lower energy state (true vacuum) through the bubble nucleation, where these states are separated from a surrounding metastable phase by an interface. The appearance of a bubble is to be considered as a regular sequence of the thermodynamical fluctuations of the system sufficiently close to a critical point of a first-order phase conversion. According to these fluctuations, bubbles are expected to create. Among them, some will grow up while others will shrink and disappear, relying on their energy competition with regard to the homogeneous preexisting metastable sate. Only the bubbles with a large size will ultimately survive and then play a decisive role in the process of nucleation of bubbles. In the end, these bubbles larger than a critical radius will expand, collide and eventually coalesce to complete a fist-order phase transition. The most popular theory of bubble nucleation was pioneered by the works of Langer in the context of classical statistical mechanics in the late 1960s \cite{Langer:1967ax,Langer:1969bc}, and it was extended by Kobzarev et al \cite{Kobzarev:1974cp} in the context of field theory. These primary jobs were put onto a solid ground in serially seminal works done by Callan and Coleman in a zero-temperature quantum field theory \cite{Coleman:1977py,Callan:1977pt,Coleman:1988}. Soon after, the studies were generalized to nonzero temperature by Affect \cite{Affleck:1980ac} and Linde \cite{Linde:1980tt,Linde:1981zj}. A main target of a nucleation theory is to determine the bubble nucleation rate per unit volume per unit time in the spirit of a semi-classical method called as the saddle-point approximation. Within this method, the key point is to find the bounce solution. The solution is actually an instanton field configuration under the extremization of the action and interpolated between the false vacuum and true vacuum between a barrier. Once the bounce is found, one can compute the crucial issues intimately related to the dynamics of a first-order phase transition, such as the critical radius, the surface tension and the saddle point action. 

Traditionally, in description of a first-order hadron quark conversion, the equation of states can be technically obtained by matching the pressures in terms of quark and hadron matters via the Maxwell construction or the Gibbs condition. The reason is that for each aspect of matters what we needed is the absolute minimum of an effective potential, which gives the pressure of the system. However, in order to calculate the bounce solution, the landscape of the whole potential is necessarily requested. In other words, a unified shape of the effective potential that contains hadron and quark phases is necessary for the existence of exact bounce solution. For a realistic nuclear matter, the nucleation rate and the surface tension have been studied and investigated in Ref.\cite{Fraga:2018cvr}, where the chiral nucleon-meson (CNM) model with the nucleon degree of freedom is employed \cite{Floerchinger:2012xd,Drews:2013hha}. The CNM model incorporated with chiral symmetry has been applied to reproduce the nuclear properties at saturation in symmetric matter. Recently, it has been utilized to study the possible paste phase at a chiral phase transition in neutron stars\cite{Schmitt:2020tac} and a first-order nuclear liquid-gas phase transition in nuclear matter \cite{Fraga:2018cvr}. Due to the lack of quark degree of freedom, the model does not include the effect of confinement and the chiral symmetry can only be treated in a very rough approximation in the spectrum of hadron states. On the contrary, with a similar chiral potential, a quark meson (QM) model can be successfully used to study the restoration of chiral symmetry at high temperature and density for two flavors \cite{Scavenius:2000qd,Donoghue:2022wrw} and three flavors \cite{Schaefer:2008hk,Chatterjee:2011jd}. In particular, within the QM model, the order parameters  for chiral symmetry restoration are well defined by quark condensates. Besides phase diagram and thermodynamics, the dynamics of a first-order chiral phase transition are also investigated in the QM model in Refs. \cite{Scavenius:2000bb,Ebert:2010eq,Pinto:2012aq,Kroff:2014qxa,Wang:2023omt}. Moreover, the model can be extended by combining dynamical quarks to the Polyakov loop fields in order to study both chiral and deconfinement phase transitions simultaneously and confront with the Lattice data directly \cite{Schaefer:2007pw,Mao:2009aq,Schaefer:2009ui}. Also, with the inclusion of the strange quark, based on the thin-wall approximation, the surface tension and phase diagram have been obtained in the three flavors PQM model in Refs. \cite{Mintz:2012mz,Stiele:2016cfs}. 

Aside from a few works in Refs. \cite{Scavenius:2000bb,Wang:2023omt}, most of studies in both QM and PQM models are applied in the thin-wall approximation, and the sea-quark contribution according to the presence of the fermion vacuum fluctuation are normally ignored in the effective potential. Because the validation of the thin-wall approximation is limited, especially when two minima of the effective potential become separated. In particular, there exists a weak first-order phase transition in phase diagram in the region of high density, where the main mechanism of a phase transition is changed to a spinodal decomposition \cite{Bessa:2008nw} as temperatures are close to the spinodal lines. Moreover, the sea quark contribution in the effective potential plays an important role in the research of QCD phase transition for the sake that it will not only soften the first-order phase transition but also decrease the surface tension of hadron-quark phases significantly. Therefore, by including the fermionic fluctuation and the decofinement, we study the dynamics of the bubble nucleation related to a first-order chiral phase transition in the PQM model from nuclear matter to quark matter and vice versa.

The remainder of present work is organized as follows. In section II, in mean field approximation, we briefly provide an overview of the Polyakov quark-meson model and present effective potential at different temperature and chemical potenital. Sect. III describes phase diagram and effective potentials at two different directions of order parameter fields. In Sec. IV, we give a brief review of the homogeneous thermal nucleation theory and build a method to search the local minima of the effective potential. Main results and discussions are supplied in Sec. V, whereas we gives our summary in the last section.

\section{Formulation of the Model}
The Lagrangian density of the QM which has an explicit realization of chiral $SU_L(2)\otimes SU_R(2)$ symmetry is given by \cite{Gell-Mann:1960mvl,Scavenius:2000qd,Donoghue:2022wrw}
\begin{equation}
{\cal L}=\overline{\psi} \left[ i\gamma ^{\mu} \partial_{\mu}-
g(\sigma +i\gamma _{5}\vec{\tau} \cdot \vec{\pi} )\right] \psi
+ \frac{1}{2} \left(\partial _{\mu}\sigma \partial ^{\mu}\sigma +
\partial _{\mu}\vec{\pi} \cdot \partial ^{\mu}\vec{\pi}\right)
-U(\sigma ,\vec{\pi}),\\
\label{Lagrangian}
\end{equation}
where the quark field $\psi$ is a flavor doublet
\begin{equation}
  \psi = \binom{u}{d}.
\end{equation}
The mesonic field $\vec{\pi} =(\pi _{1},\pi _{2},\pi _{3})$ is a pseudoscalar pion fields while $\sigma$ is a spin-$0$ scalar field. $g$ is a flavor-blind Yukawa interaction coupled with quarks and mesons.

The pure mesonic potential, which exhibits both spontaneously and explicitly broken chiral symmetry, is given by the expression
\begin{equation}
U(\sigma ,\vec{\pi})=\frac{\lambda}{4} \left(\sigma ^{2}+\vec{\pi} ^{2}
-{\vartheta}^{2}\right)^{2}-H\sigma.
\label{mpot}
\end{equation}
Here $\lambda$ is a positive coupling constant for the mesonic fields, the constant $\vartheta$ is a physical quantity called the vacuum expectation value of scalar field. The last term in above equation represents finite masses for current quarks, which is decided by the partially conserved axial-vector current (PCAC) relationship with the expression as $H=f_{\pi}m_{\pi}^{2}$.

Besides the chiral symmetry, it is well-known that the basic strong-interaction theory QCD has another salient feature, the confinement. The confinement is related to the center symmetry of the color gauge group and the center symmetry is believed to get spontaneously broken in the high energy region of quark matter. In order to incorporate the physical aspect of the confinement-deconfinement phase transition in the QM model, the Polyakov loop operator that is related to the free energy of a static color charge is usually introduced and applied. In such an extension, the Polyakov loop operator is introduced as a path ordered Wilson loop of the gauge field in temporal direction,
\begin{eqnarray}
L(\vec{x})=\mathcal{P}\mathrm{exp}\left[i\int_0^{\beta}d \tau
A_4(\vec{x},\tau)\right],
\end{eqnarray}
where $\mathcal{P}$ denotes path ordering, $\beta=1/T$ being the inverse of temperature and $A_4=iA^0$. Accordingly, the color traced Polyakov loops are set as
\begin{eqnarray}
\Phi=(\mathrm{Tr}_c L)/N_c, \qquad \Phi^*=(\mathrm{Tr}_c
L^{\dag})/N_c.
\end{eqnarray}
Accordingly, the thermal expectation values $\langle \Phi \rangle$ and $\langle \Phi^* \rangle$ are to be defined as the Polyakov loop variables. But for simplicity, we still use $\Phi$ and $\Phi^*$ for their arguments. Furthermore, $\Phi$ and $\Phi^*$ are complex scalar fields if we take a diagonal representation as shown in Refs. \cite{Fukushima:2003fw,Fukushima:2017csk}. In a pure gauge theory, the Polyakov loop variables offers an order parameter for deconfinement phase transition. It is to vanish in the confined hadron matter where the quarks receive infinite free energies. On the contrary, it becomes finite in the deconfined quark matter. However, in a gauge theory with dynamical quarks, it should be taken as a sketchy order parameter only. The situation is roughly analogous with the chiral condensate which is an exact order parameter for massless quark in chiral limit, whereas it is only a rough order parameter for massive quarks.

When the Polyakov loop has been included in the QM model, the dynamical quarks is now coupled with gauge fields through the following covariant derivative: $D_{\mu}=\partial_{\mu}-i A_{\mu}$ and $A_{\mu}=\delta_{\mu 0}A_0$. In mean-field approximation, both meson and gauge fields are taken as the classical fields and their thermal and quantum fluctuations are neglected, whereas the fermionic quarks are treated as dynamical fields. After integration over the fermions, the grand thermodynamical potential density in the presence of the Polyakov loop can then be written as
\begin{eqnarray}\label{omegapsi}
\Omega_{\bar{\psi} \psi} =\Omega_{\bar{\psi} \psi} ^{\mathrm{vac}}+\Omega_{\bar{\psi} \psi}^{\mathrm{th}} =-2N_f N_c\int \frac{d^3\vec{p}}{(2
\pi)^3}E_q  -2 N_f T \int \frac{d^3\vec{p}}{(2
\pi)^3} \left[ \mathrm{ln} g_q^+ + \mathrm{ln} g_q^- \right].
\end{eqnarray}
Here, $N_f=2$, $N_c=3$, $E_q=\sqrt{\vec{p}^2+m_q^2}$, and the constituent quark masses for $u$ and $d$ quarks are given by
$m_q=g \sigma_v$ together with $\sigma_v\equiv \langle \sigma\rangle$. In vacuum $\sigma_v=f_{\pi}=93.0 \mathrm{MeV}$. The first term of Eq.(\ref{omegapsi}) represents the one-loop fermion vacuum contribution, which is divergent and should be renormalized by using a regularization scheme. In order to obtain the physical quantities independent of the renormalization scale, in what follows, we would like to take a more convenient scheme by using the the dimensional regularization. The second term of Eq.(\ref{omegapsi}) $g_q^+$ and $g_q^-$ have been defined upon taking trace over color space
\begin{eqnarray}
 g_q^+ &=& \left[ 1+3(\Phi+\Phi^*e^{-(E_q-\mu)/T})\times e^{-(E_q-\mu)/T}+e^{-3(E_q-\mu)/T} \right],  \\
 g_q^- &=& \left[ 1+3(\Phi^*+\Phi e^{-(E_q+\mu)/T})\times e^{-(E_q+\mu)/T}+e^{-3(E_q+\mu)/T} \right].
\end{eqnarray}

The first term of Eq.(\ref{omegapsi}) is just the fermion vacuum one-loop contribution to the effective potential at $T=\mu=0$ and the integral is ultraviolet divergent. To regularize this fermion vacuum term, we need to perform the dimensional regularization near three dimensions, $d=3-2\epsilon$, in order to isolate the divergences. This yields a resulting potential up to zeroth order in $\epsilon$ as given by \cite{Quiros:1999jp,Laine:2016hma}
\begin{eqnarray}\label{vacterm}
\Omega_{\bar{\psi} \psi} ^{\mathrm{vac}}=\frac{N_f N_c}{16 \pi^2}m^4_q \left[\frac{1}{\epsilon}+\ln\frac{\Lambda^2}{m_q^2}+\ln(4\pi)-\gamma_E+\frac{3}{2}\right],
\end{eqnarray}
where $\Lambda$ is an arbitrary renormalization scale parameter. Then the effective potential can be renormalized by adding a proper counter term to the Lagrangian as done for the two-flavor case in Refs. \cite{Skokov:2010sf,Gupta:2011ez}. Up to some irrelevant constants, the renormalized fermion vacuum loop contribution reads
\begin{eqnarray}\label{omegareg}
\Omega_{\bar{\psi} \psi} ^{\mathrm{vac}}=\Omega_{\bar{\psi} \psi} ^{\mathrm{reg}}=-\frac{N_c N_f}{8\pi^2} m_q^4 \mathrm{ln}(\frac{m_q}{\Lambda}),
\end{eqnarray}
It is worth pointing out that the parameters $\lambda$ and $\vartheta$ are technically dependent on the arbitrary renormalization scale $\Lambda$, however, as indicated in Refs. \cite{Jin:2015goa,Skokov:2010sf,Gupta:2011ez,Li:2018rfu}, the $\Lambda$ dependence can be neatly cancelled out by redefining the parameters in the model, and the physical quantities in the present model are therefore free from ambiguities related to the choice of the renormalization scale.

The phenomenological potential of the Polyakov loop $\mathbf{\mathcal{U}}(\Phi,\Phi^*,T)$ as a function of $T$, $\Phi$ and $\Phi^*$ is constructed to duplicate some physical quantities of the pure gauge theory calculated on lattice QCD \cite{Svetitsky:1985ye,Ratti:2022qgf}. The first one is that the terms of the potential $\mathbf{\mathcal{U}}$ must be invariant under the $Z(3)$ center symmetry since the pure gauge QCD theory conserves this symmetry. Secondly, for a pure gauge theory, because there is no fundamental principle to give out any asymmetry between $\Phi$ and $\Phi^*$, the potential $\mathbf{\mathcal{U}}$ should have additional requirement with the symmetry under the exchange between $\Phi$ and $\Phi^*$. Finally, as mentioned as above, $\Phi$ and $\Phi^*$ are treated as order parameters for the deconfinement phase transition, we thus require that the thermal expectation values $\Phi$ and $\Phi^*$ evaluated at the minimum of the potential $\mathbf{\mathcal{U}}$ are $\Phi=\Phi^*=0$ when temperature is low, on the contrary, when the temperature is high, these values should become finite and asymptotically approach to a constant, e.g. the unity $1$ for the logarithmic form of the Polyakov-loop potential. For the specific functional form of the potential $\mathbf{\mathcal{U}}$, there still exist various possibilities in the literature with similar properties \cite{Ratti:2005jh,Roessner:2006xn,Schaefer:2007pw,Fukushima:2008wg,Schaefer:2009ui}. For the physical interest in the present work, we prefer to use the simplest polynomial potential based on a Ginzburg-Landau ansatz as suggested in Ref.\cite{Ratti:2005jh}
\begin{eqnarray}\label{polpential}
\frac{\mathbf{\mathcal{U}}(\Phi,\Phi^*,T)}{T^4}=-\frac{b_2(T)}{4}(|\Phi|^2+|\Phi^*|^2)-\frac{b_3
}{6}(\Phi^3+\Phi^{*3})+\frac{b_4}{16}(|\Phi|^2+|\Phi^*|^2)^2,
\end{eqnarray}
with a temperature dependent coefficient
\begin{eqnarray}
b_2(T)=a_0+a_1\left(\frac{T_0}{T}\right)+a_2\left(\frac{T_0}{T}\right)^2+a_3\left(\frac{T_0}{T}\right)^3.
\end{eqnarray}
The parameters in above potential are adopted to confront with the lattice results for the pure gauge theory. This yields the following values
\begin{eqnarray}
a_0=6.75,\qquad a_1=-1.95,\qquad a_2=2.625, \nonumber \\
 a_3=-7.44,\qquad
b_3=0.75,\qquad b_4=7.5.
\end{eqnarray}
Moreover, for pure $SU(3)$ gauge theory, there is a first-order deconfinement phase transition with a critical temperature at $T_0=270$ MeV. However, by taking account of the back action of the dynamical quarks on the gluonic sector, it is natural to let $T_0=T_0(N_f)$ in order to guarantee that the behavior in the glue sector also depends on the number of quark flavors or even the baryon chemical potential \cite{Schaefer:2007pw,Haas:2013qwp}. In the following study we shall use $T_0=208$ MeV for two quark flavors \cite{Gupta:2011ez,Schaefer:2007pw}.

Now the thermodynamic grand potential for the PQM model in the mean-field approximation with the fermion vacuum one-loop is given by
\begin{eqnarray}\label{potential}
\Omega_{\mathrm{MF}}(T,\mu)=U(\sigma ,\vec{\pi} )+\mathbf{\mathcal{U}}(\Phi,\Phi^*,T)+\Omega_{\bar{\psi}
\psi}.
\end{eqnarray}
The equations of motion for the mesonic and gluonic fields $\sigma$, $\Phi$ and $\Phi^*$ can be derived by the stationarity conditions
\begin{eqnarray}\label{gapequs}
\frac{\partial \Omega_{\mathrm{MF}}}{\partial \sigma_v}=0, \qquad \frac{\partial \Omega_{\mathrm{MF}}}{\partial
\Phi}=0, \qquad \frac{\partial \Omega_{\mathrm{MF}}}{\partial \Phi^*}=0,
\end{eqnarray}
which yield the order parameters $\sigma(T, \mu)$, $\Phi(T, \mu)$ and $\Phi^*(T, \mu)$ as a function of temperature and chemical potential. The parameters in the model are specified by their vacuum properties and their physical values are set at $m_{\pi}=138$ MeV, $m_{\sigma}=500$ MeV and $f_{\pi}=93$ MeV. For the last coupling constant $g$, it is usually determined by the constituent quark mass in vacuum as $g=3.3$.

\section{phase structure}

\begin{figure}
\includegraphics[scale=0.36]{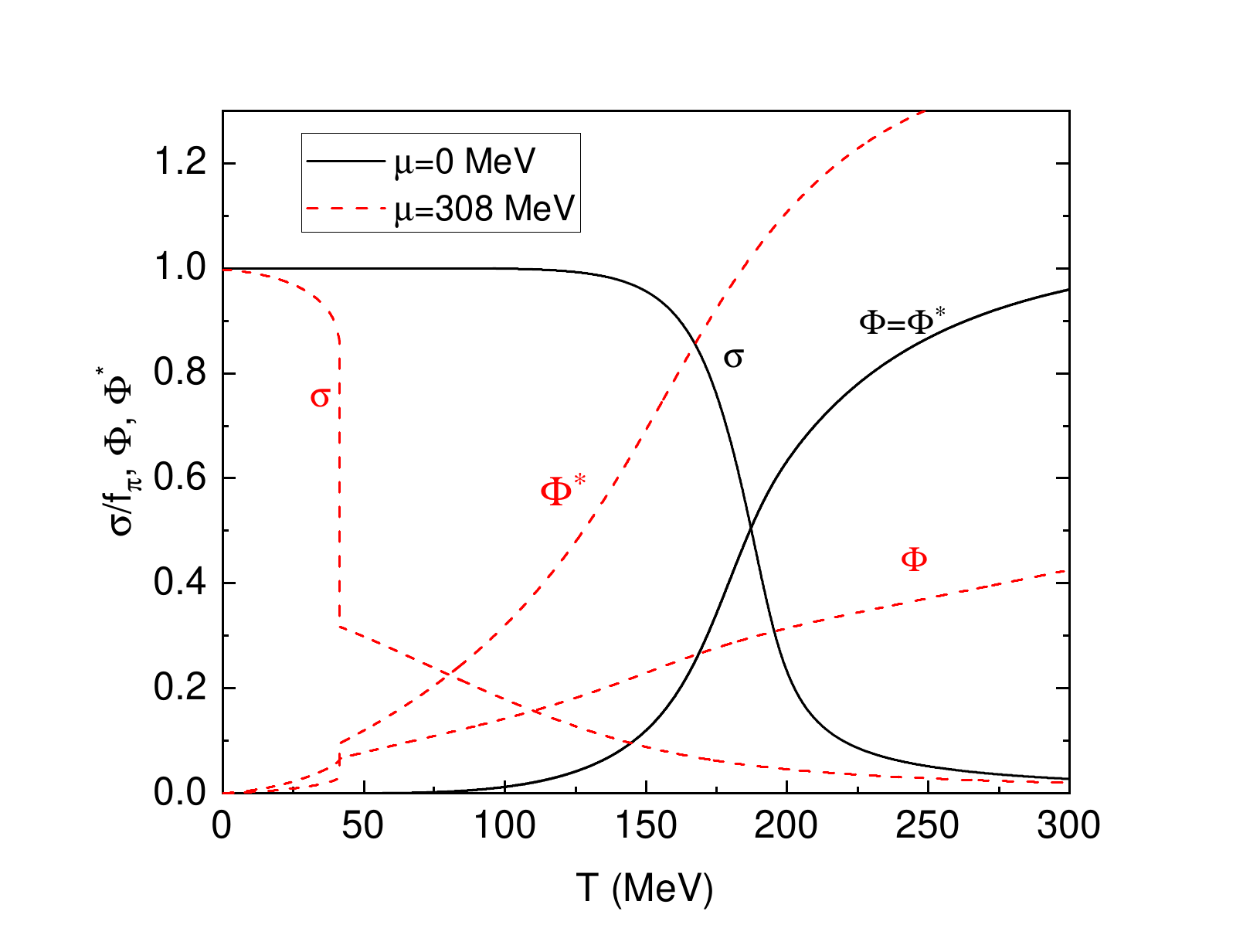}
\caption{\label{Fig01} (Color online) The normalized chiral order parameter $\sigma$ and the Polyakov loop expectation values $\Phi$, $\Phi^*$ as functions of temperature for $\mu = 0$ MeV and $\mu = 308$ MeV. The set of solid curves are for  $\mu = 0$ MeV and the dashed curves are for $\mu = 308$ MeV.}
\end{figure}

After solving the above coupled gap equations (\ref{gapequs}), we can obtain the chiral condensate $\sigma$ and the Polyakov loop expectation values $\Phi$ and $\Phi^*$ as a function of the temperature for different chemical potential. The temperature behaviors of these order parameters could be used to reveal the nature of chiral and deconfinement phase transition. For demonstration, the temperature variations of the chiral condensate and Polyakov loop expectation values are displayed in Fig.\ref{Fig01} when the chemical potentials are set to $\mu = 0$ MeV and $\mu = 308$ MeV. In order to illustrate the subtle differences between these two types of QCD phase transition, we would like to split our studies into two parts, one is the chiral phase transition related to the chiral symmetry, another is the deconfinement phase transition that is directly relevant to the $Z(3)$ center symmetry.

Let us first study the restoration of chiral symmetry at various chemical potentials. The chiral condensate as a function of temperature shows that the system experiences a quite smooth crossover conversion at zero chemical potential. Conversely, for a relatively larger chemical potential, i.g. $\mu=308$ MeV, the order of phase transition has been changed to a first-order since the chiral condensate makes a sharp jump across the gap of the condensate when the temperature is nearby its critical value. Therefore, by compared two curves of the chiral condensate $\sigma$ in Fig.\ref{Fig01}, we find that the chemical potential would make the variation of chiral order parameter with temperatures sharp and sharper. Subsequently, the crossover will eventually turn into a second-order phase transition at somewhere called as a CEP, where the order phase conversion is of second. The appearance of the CEP guarantees the existence of a first-order chiral phase transition, and vice versa. From Fig.\ref{Fig01}, for both kind of phase transitions, the temperature derivative of the chiral condensate $\sigma$ has merely one typical peak at a peculiar temperatures, which is usually defined as the critical temperature for chiral phase transition. For vanishing chemical potential, the chiral restoration happens around $T_{\chi}^c \simeq 187$ MeV, meanwhile for a larger chemical potential at $\mu=308$ MeV, the critical temperature of chiral phase transition will go down to the lower temperature about $T_{\chi}^c \simeq 41$ MeV.

This is unlike the chiral phase transition, in which the critical temperature is well defined at least for the case of a first-order phase transition. On the contrary, we are still ambiguous to safely confirm the deconfinement phase transition through the order parameters $\Phi$, $\Phi^*$ or their temperature derivatives \cite{Fukushima:2008wg,Mao:2009aq}, even though the chemical potential is very large. In order to explicate this problem, the temperature variations of the Polyakov loop expectation values $\Phi$ and $\Phi^*$ at $\mu = 0$ MeV and $\mu = 308$ MeV are demonstrated in Fig.\ref{Fig01}. From this figure, by taking the $\Phi$ as example, we can find that the evolution of the Polyakov loop expectation value $\Phi$ as a function of the temperature exhibits a rather smooth behaviour for zero chemical potential. For a larger chemical potential at $\mu = 308$ MeV, $\Phi$ becomes more smoother and flatter except that there is a small jump coincident with the jump of the chiral order parameter $\sigma$. Because the temperature derivative $\Phi'=d \Phi/dT$ generally has one more peak in calculations, especially when $\mu$ is high \cite{Gupta:2011ez,Mao:2009aq}, it is obscure to determine the critical temperature for the deconfinement phase transition through the peaks of the Polyakov loop variation. The problem will arise again and become worse for chiral and deconfinement phase transition when the strange quark $s$ is included in POM model \cite{Schaefer:2009ui,Mao:2009aq}.

In the following study, we will use the effective potential to reveal more information about the natural properties of the chiral and deconfinement phase transitions at finite temperature and chemical potential. Besides solving the equations of motion (\ref{gapequs}), based on the shape or geometry of an effective potential, we can directly search a global minimum of the effective potential $\Omega_{\mathrm{MF}}$ in Eq.(\ref{potential}) to get the expectation values of the chiral condensate and the Polyakov loop fields, or more interestingly, the local minima of the effective potential, which are corresponding to the false vacua during phase transitions. For convenience, we refer a study based on the geometry of effective potential as a geometric approach. For a single order parameter, such as the QM model, a geometric approach can precisely tell two types of chiral phase transition and simultaneously give out two minima of effective potential\cite{Scavenius:2000qd,Mao:2019aps}. However, for the PQM model there are three order parameter variables $\sigma$, $\Phi$ and $\Phi^*$ in the grand canonical potential $\Omega_{\mathrm{MF}}$ in Eq.(\ref{potential}), so that it is extremely challenging to demonstrate and reveal the thermal effective potential via evolving these variables simultaneously in such a large number. Inspired by previous studies in the PQM model \cite{Jin:2015goa,Li:2018rfu}, in order to simplify the problem and provide a more intuitive insight into the physics, the number of the active variable in the effective potential at finite temperature and chemical potential is to be set as one at a time. This means that the effective potential has been divided into two directions: (1) the gluonic direction, and (2) the mesonic direction. In the gluonic direction, the Polyakov loop $\Phi$ (or $\Phi^*$) is considered as the active variable while the $\sigma$ and $\Phi^*$ (or $\Phi$) fields remain on their expectation values all the time. By contrast, in the mesonic direction, the $\sigma$ field is taken as a active variable in the grand canonical potential $\Omega_{\mathrm{MF}}$ while the Polyakov loop fields are set on their expectation values $\Phi$ and $\Phi^*$.

\begin{figure}[thbp]
\epsfxsize=9.0 cm \epsfysize=6.5cm
\epsfbox{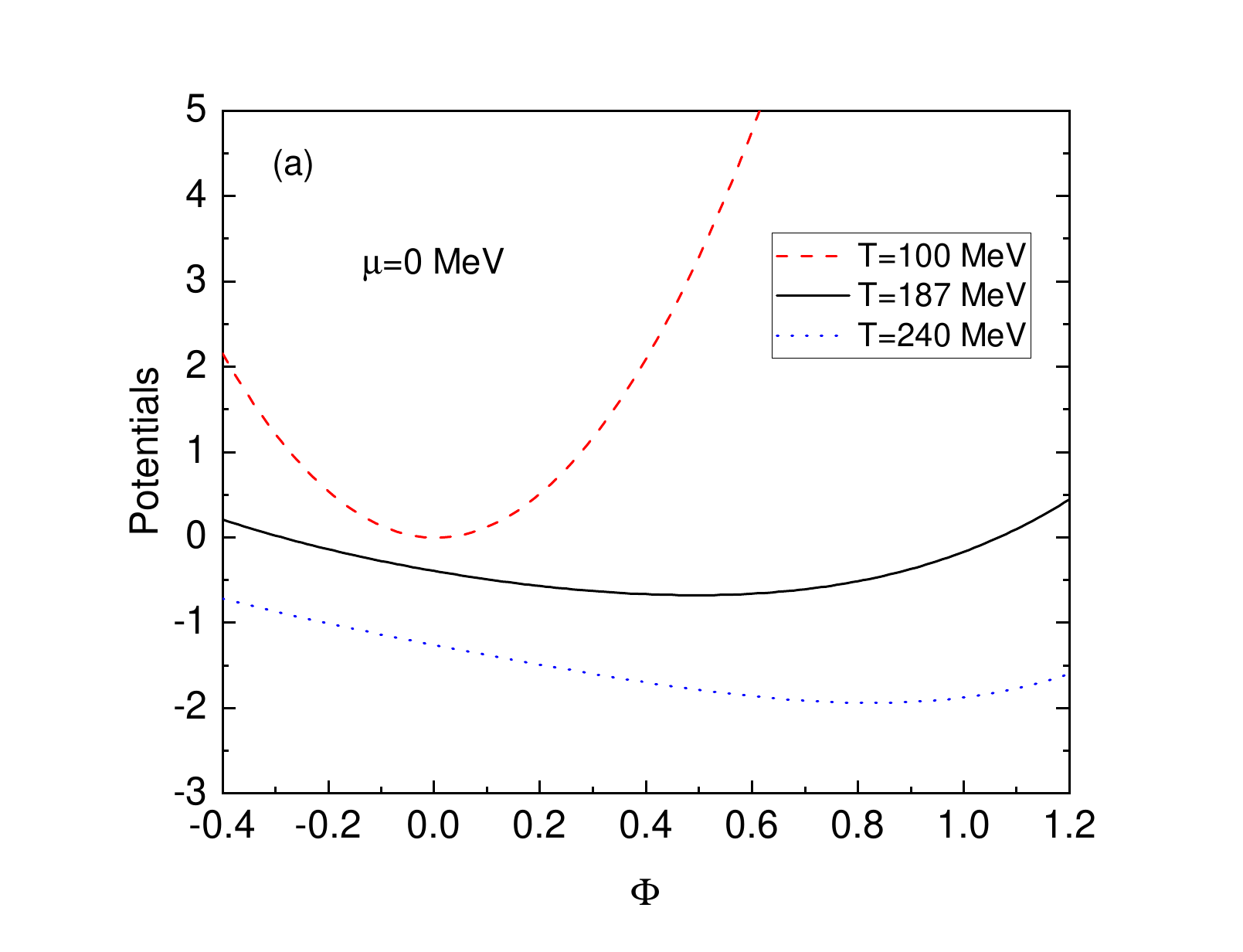}\hspace*{0.01cm} \epsfxsize=9.0 cm
\epsfysize=6.5cm \epsfbox{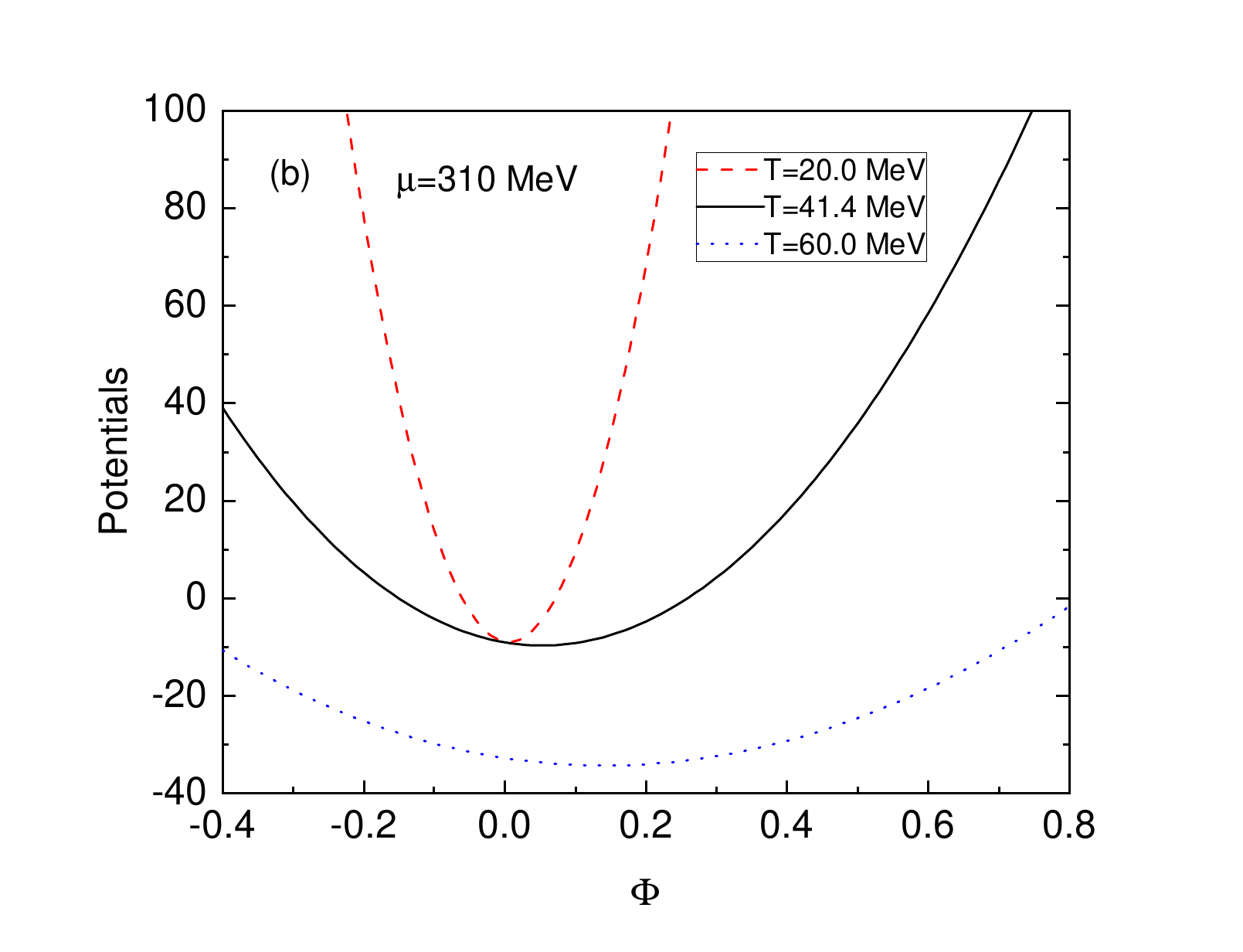}
 \caption{(Color online) (a) The scaled grand canonical potentials $\Omega_{\mathrm{MF}}$ over $T^4$ as a function of the Polyakov loop $\Phi$ for $\mu=0$ MeV by fixing the $\sigma$ and $\Phi^*$ fields on their expectation values. (b)The scaled grand canonical potentials $\Omega_{\mathrm{MF}}$ over $T^4$ as a function of the Polyakov loop $\Phi$ for $\mu=310$ MeV by fixing the $\sigma$ and $\Phi^*$ fields on their expectation values. $\Omega_{\mathrm{MF}}$ is scaled by a factor of $T^4$.}
\label{Fig02}
\end{figure}

For simplicity, we firstly take the Polyakov loop $\Phi$ as a typical example in the following discussion, the qualitative result presented here, of course, can be applied to that of the $\Phi^*$ field too. The gluonic case is shown in Fig.\ref{Fig02}, where the left panel is the scaled grand canonical potentials $\Omega_{\mathrm{MF}}$ over $T^4$ as a function of the Polyakov loop $\Phi$ for zero chemical potential, and the right panel is the scaled grand canonical potentials $\Omega_{\mathrm{MF}}$ over $T^4$ as a function of the Polyakov loop $\Phi$ for $\mu=310$ MeV. From Fig.\ref{Fig02} it is obvious that the grand canonical potentials for both $\mu=0$ MeV and $\mu=310 $ MeV share similar behaviors: there is only one minimum for each of the grand canonical potentials. These minima, which are to be identified as the expectation values $\Phi$ of the Polyakov loop field, move to their higher values smoothly and slowly with the increase of the temperature. Even though the chemical potential is so large as $\mu=310$ MeV, when the chiral phase transition is of first-order, the grand canonical potential $\Omega_{\mathrm{MF}}$ still exhibits a trivial feature with a single minimum. This indicates that the disconnection of the order parameter $\Phi$ in Fig.\ref{Fig01} for $\mu=308$ MeV and $T \simeq 41$ MeV should be driven by that of the chiral order parameter $\sigma$, such a sudden jump does not supported by the grand canonical potential of the Polyakov loop field itself. So that the peak in the Polyakov loop field derivative around the critical temperature at $T=T_{\chi}^c \simeq 41$ MeV is a fake signal for the definition of the deconfinement phase transition. This is also the reason why a critical temperature of the deconfinement conversion should be given by the another pseudo peak corresponding to $\Phi(T)/\Phi(T\rightarrow \infty)>1/2$ as discussed in Refs. \cite{Gupta:2011ez,Jin:2015goa}.

Theoretically, the trivial behavior of the grand canonical potential as a function of the Polyakov loop can be interpreted qualitatively if we go back to the original pure-gluonic potential $\mathbf{\mathcal{U}}$ in Eq. (\ref{polpential}). It is easy to show that the potential $\mathbf{\mathcal{U}}$ displays only one minimum at $\Phi=0$ as long as the temperature is less than the critical temperature at $T_0=208$ MeV. Only when $T$ is larger than the critical value $T_0$, the potential $\mathbf{\mathcal{U}}$ will develop a typical first-order potential with two local minima separated by a global maximum (or a barrier). It seems that the inclusion of the quark and meson fields in the grand canonical potential $\Omega_{\mathrm{MF}}$ does not change the nature of the deconfinement phase dramatically. As demonstrated in Fig.\ref{Fig02}, with the increase of the temperature, the global minimum of the potential moves continuously from its low value to a high one, but the characteristic shape of the potential remains intact as long as $T<T_0$.   

\begin{figure}[thbp]
\epsfxsize=9.0 cm \epsfysize=6.5cm
\epsfbox{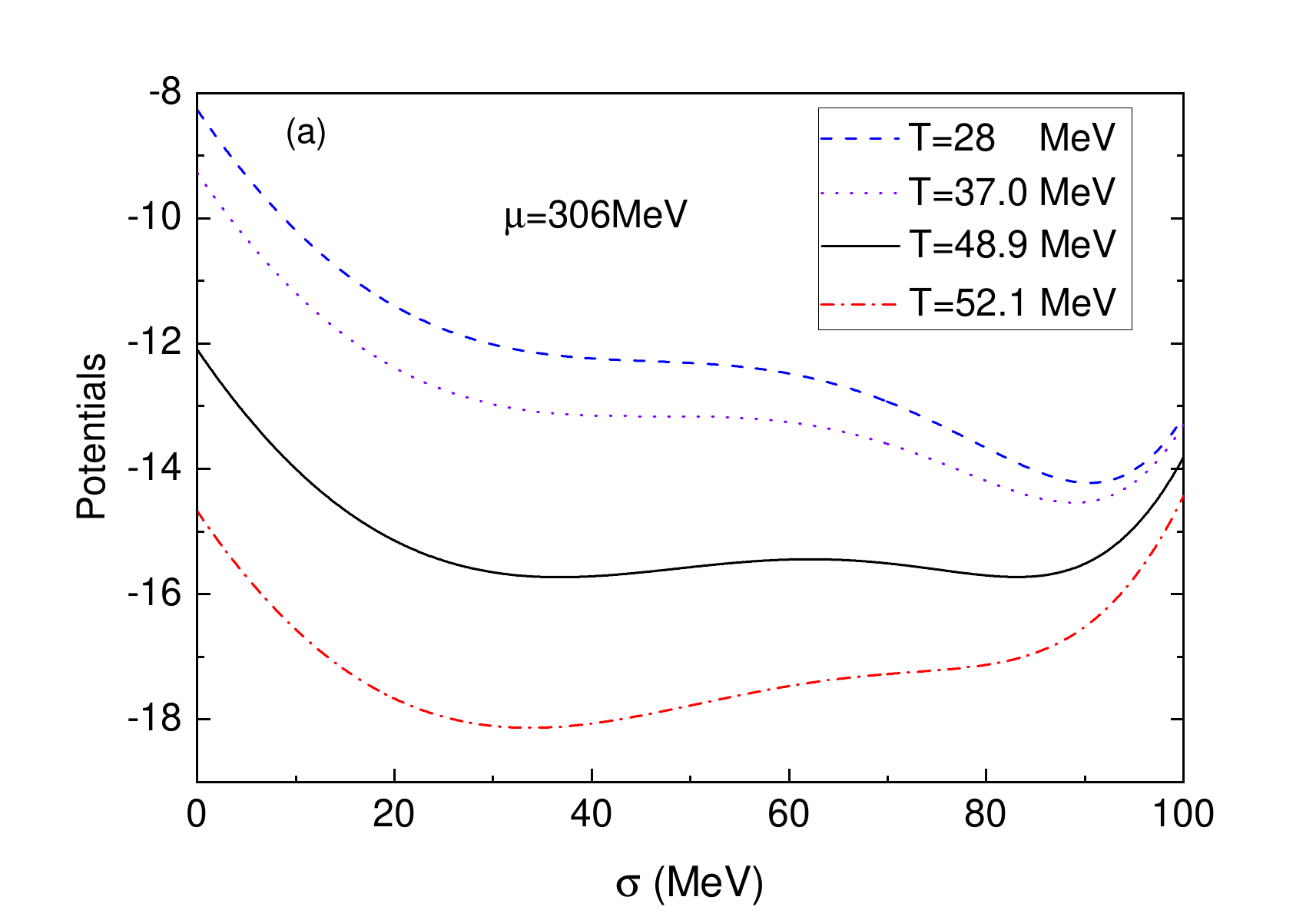}\hspace*{0.01cm} \epsfxsize=9.0 cm
\epsfysize=6.5cm \epsfbox{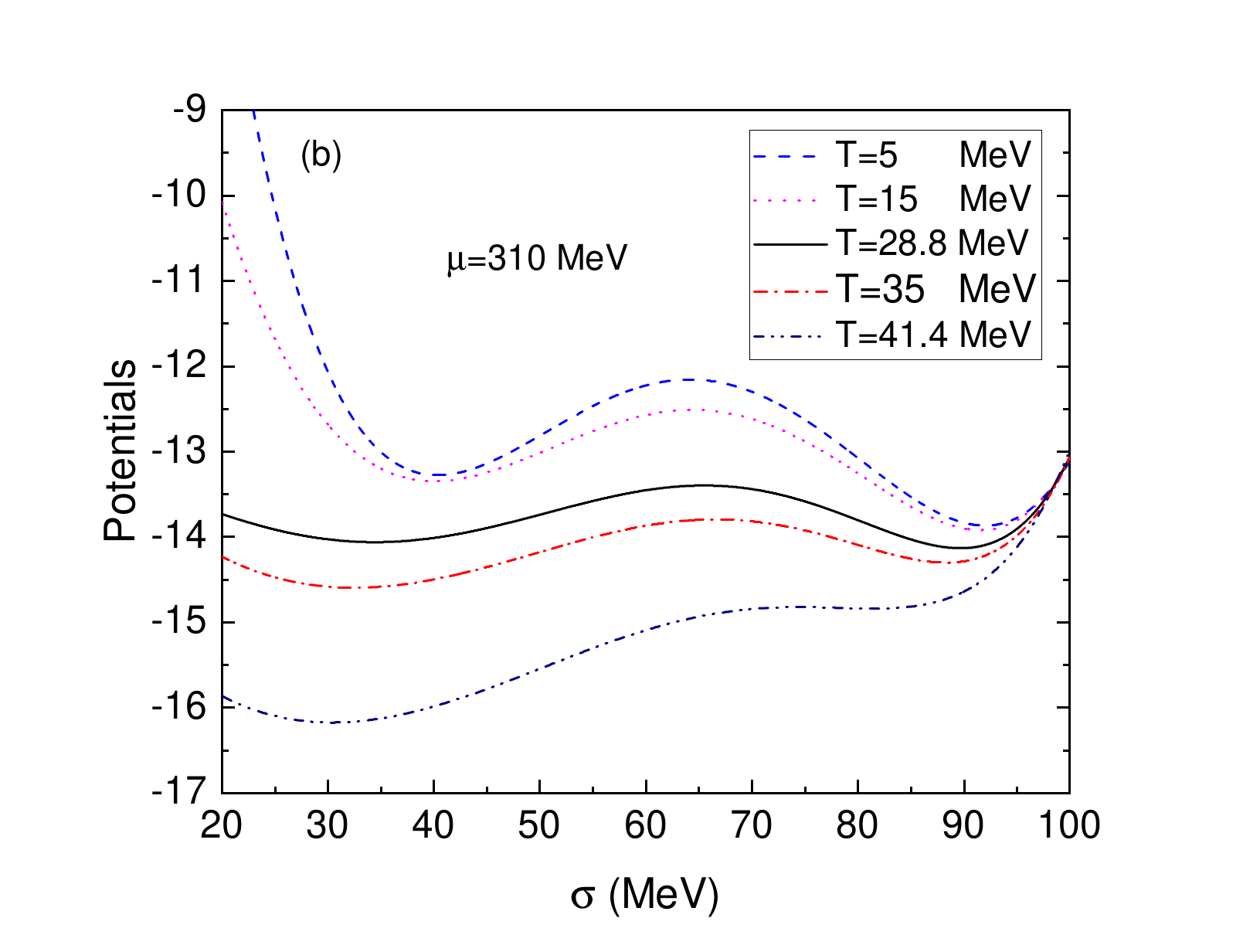}
 \caption{(Color online) (a) The grand canonical potentials $\Omega_{\mathrm{MF}}$ as a function of the chiral order parameter $\sigma$ for $\mu=306$ MeV by fixing the Polyakov loop on their expectation values. (b)The grand canonical potentials $\Omega_{\mathrm{MF}}$ as a function of the chiral order parameter $\sigma$ for $\mu=310$ MeV by fixing the Polyakov loop on their expectation values. The unit of the effective potential $\Omega_{\mathrm{MF}}$ is scaled as $\mathrm{MeV/fm^3}$.}
\label{Fig03}
\end{figure}

Let us now investigate how the grand canonical potentials $\Omega_{\mathrm{MF}}$ evolving with the chiral order parameter $\sigma$ for different chemical potentials by fixing the Polyakov loops on their expectation values. The scaled grand canonical potentials are shown in Fig.\ref{Fig03} as a function of $\sigma$ for $\mu=306$ MeV and $\mu=310$ MeV, respectively. From Fig.\ref{Fig03}, one clearly observes the characteristic pattern of a first-order phase transition potential when $\mu$ is very large: two minima corresponding to phases of restored and broken chiral symmetry are separated by a potential barrier and they will become degenerate at $T=T_{\chi}^c$. Chiral symmetry is approximately restored as $T>T_{\chi}^c$, where the previous local minimum at relatively low $\sigma$ value becomes the global/absolute minimum as shown in Fig.\ref{Fig03}. On the other hand, when the temperature is below the critical one $T_{\chi}^c$, the global minimum locates in a relatively larger sigma and the constituent quark becomes massive. In this time, the chiral symmetry is spontaneously broken. As a result, we can conclude that there is a firs-order chiral phase transition when chemical potentials are both at $\mu=306$ MeV and $\mu=310$ MeV. Furthermore, when the temperature is at $T=T_{\chi}^c$, the height of the potential barrier is about $0.66$ $\mathrm{MeV/fm^3}$ as $\mu=310$ MeV, but it will drop sharply to $0.30$ $\mathrm{MeV/fm^3}$ when the chemical potential reduces to $\mu=306$ MeV. So that with decreasing of the chemical potential, it is believed that the potential barrier combined two minima becomes flat and flatter. Consequently, the first order chiral phase transition turns into a second order phase transition at a specific point called as the CEP. In that moment the barrier disappears and the chiral phase transition exhibits a second-order quality.

For a large chemical potential, the grand canonical potentials $\Omega_{\mathrm{MF}}$ as a function of the chiral order parameter $\sigma$ in Fig.\ref{Fig03} really guarantee the existence of the first-order chiral phase transition. But the nature of the phase transition revealed by the shapes of effective potentials is fairly distinct from each other. For $T>T_{\chi}^c$, both cases show that there have the critical spinodal temperatures $T_{\mathrm{sp}}$ since the local minima of effective potential are going to disappear at certain temperature. This suggests that the hadron-quark phase transition is a weak first-order phase transition if the system is beating up from low temperature to a high one. However, while the temperature $T$ is below the critical value $T_{\chi}^c$, the evolutions of the potential with the order parameter $\sigma$ for these two chemical potentials exhibit quite different features. From the left panel in Fig.\ref{Fig03}, for temperatures very near the critical temperature $T_{\chi}^c \simeq 48.9$ MeV, the grand canonical potentials $\Omega_{\mathrm{MF}}$ for $\mu=306$ MeV indeed displays a local minimum which is separated by a barrier from another local minimum. But as the temperature is lowered, the local minimum at low $\sigma$ value approaches the intervening maximum. These two minima of the potential meet and form an inflection point at the spinodal temperature $T_{\mathrm{sp}}\simeq 37$ MeV. Accordingly, there has only one minimum in the grand canonical potentials $\Omega_{\mathrm{MF}}$ when $T\leq T_{\mathrm{sp}}$. Because of the disappearance of the barrier between two local minima in the effective potential, this kind of the phase transition sometime can be denoted as a weak first-order conversion \cite{Wang:2023omt,Bessa:2008nw}. On the contrary, the increase of the chemical potential normally leads to a larger barrier between the two degenerate minima of the grand canonical potentials $\Omega_{\mathrm{MF}}$ when $T$ is close to $T_{\chi}^c$. In other words, the first order phase transition would be strengthened with the growth of the chemical potential, and a larger barrier implies a larger latent heat. For the case of $\mu=310$ MeV, the grand canonical potentials $\Omega_{\mathrm{MF}}$ as a function of the chiral order parameter $\sigma$ for different temperatures are shown in the right panel in Fig.\ref{Fig03}. One significant difference from the case of $\mu=306$ MeV is that there always exists a potential barrier between two minima of the potential, no matter what the temperature is chosen. This indicates that the chiral phase transition for a quark-hadron phase transition is now a strong first-order phase transition when $\mu=310$ MeV \cite{Wang:2023omt,Zhou:2020bzk}.

\begin{figure}
\includegraphics[scale=0.36]{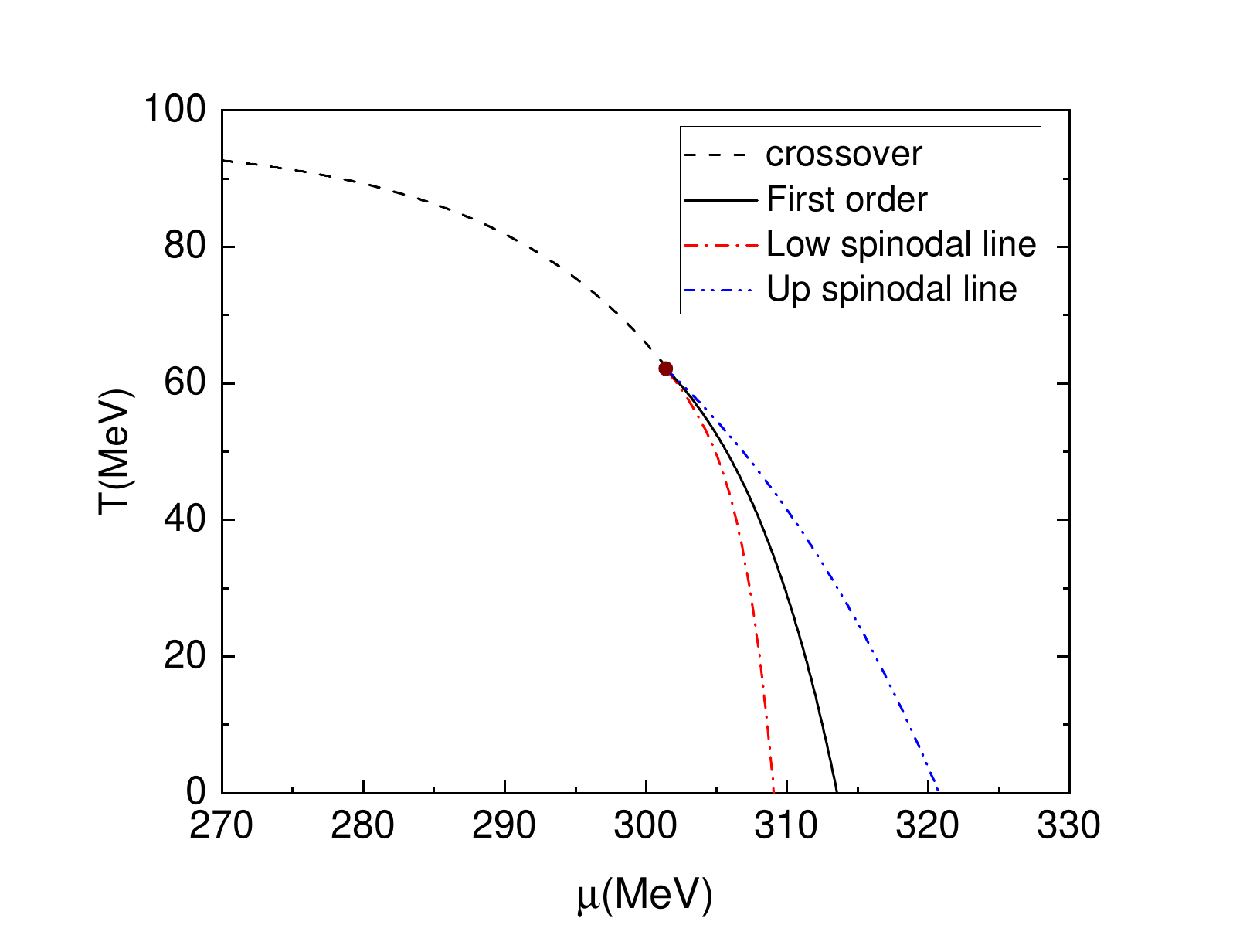}
\caption{\label{Fig04}(Color online) Phase diagram in the $T-\mu$ plane in the PQM for the chiral phase transition. The dashed curve denotes the critical line for the crossover transition. The solid curve indicates the first-order phase transitions. The solid circle represents the CEP for chiral phase transitions of $u$ and $d$ quarks, while The dashed-dotted curve and the dashed-doted-dotted curve are the lower and upper spinodal lines.}
\end{figure}

The chiral phase transition of $u$ and $d$ quarks in the $T-\mu$ plane is displayed in Fig.\ref{Fig04}. It is found that for two light flavors, there are crossover in low chemical potential and first-order phase transition in high chemical potential, and the CEP is located at $(T_{\mathrm{E}},\mu_{\mathrm{E}})\simeq(301.4 \mathrm{MeV}, 62.1 \mathrm{MeV})$. In the $T-\mu$ plane, a vast part of the QCD phase diagram is a crossover, only a very small window is left for the first order phase transition. As compared with previous studies in the QM model, the incorporation of effects of deconfinement in the language of the Polyakov-loop fields merely increase the critical temperature for chiral phase transition, while leaves the nature of the chiral phase transition unchanged. Since we are interested in the dynamics of the first order phase transition in the present study, the lines for the deconfinement crossover transitions are omitted in Fig.\ref{Fig04}. 

In order to present a more detailed description about the chiral first-order phase transition, two additional spinodal lines are also included in Fig.\ref{Fig04}. These two spinodal lines are usually introduced to characterize the region of spinodal instability for a weak first-order phase transition. From the figure \ref{Fig04}, both the lower and upper spinodal lines rise up with the reduction of the chemical potential, but the distance between these two lines become small and small. In the end, the spinodal lines together with the coexistent critical line will bend forward to the CEP and terminate at that point. Furthermore, according to the low spinodal line as $T<T_{\chi}^c$, the structure of phase diagram below the coexistent critical line can be further classified into two categories: a week first-order phase transition and a strong one. For a weak first-order phase transition, it is usually characterized by the low barrier and the existence of the critical spinodal temperature $T_{\mathrm{sp}}$, whereas for a strong first-order phase transition, there has an effective potential with a zero-temperature potential barrier. Therefore, as shown in Fig.\ref{Fig04}, along horizontal axis, the corresponding chemical potential, when a week first-order phase transition turns into a strong one, is established as $\mu_{\mathrm{c}}=309$ MeV. For $\mu>\mu_{\mathrm{c}}$, it is a strong first-order phase transition and a potential barrier separated two minima never disappears with the decrease of the temperature, conversely, the local minimum aside from the global minimum will gradually disappear and there is no potential barrier for $T<T_{\mathrm{sp}}$. On the other side of the coexistent critical line, since there exists the critical spinodal temperature, the hadron-quark phase conversion should be always of weakly first-order. 

\section{Nucleation in the PQM model}

\subsection{Homogeneous thermal nucleation}

For a first-order phase transition, the distinguished symbol of a potential is an energy barrier between two minima. For some potentials a barrier already exists at the classical tree-level, for others it could be induced by loop expansions or the integration of other dynamical fields. When the temperature is very close to its critical value $T_c$, these two minima are degenerate. However, as the temperature is deviating from $T_c$, one of the minimum becomes an absolute/global minimum (a true vacuum), the other is now changed to a local minimum (a false vacuum). The decay of a false vacuum resides in a first-order phase transition and the conversion may take place through the bubble nucleation by quantum or thermal fluctuations, relying on the ambient temperature in comparison with the barrier. In the present work, we assume a limit that thermal fluctuations dominate quantum fluctuations.

Based on the theory of homogeneous thermal nucleation, at finite temperature, we start with a four-dimensional Euclidean action for a scalar order parameter field, e.g. a scalar field $\phi$,
\begin{equation}\label{Euaction}
S_E[\phi]=\int_0^{\beta}d \tau \int dr^3 \left[ \frac{1}{2} \left(\frac{\partial \phi}{\partial \tau} \right)^2+\frac{1}{2} \left(\nabla\phi \right)^2+V(\phi) \right],
\end{equation}
where the potential has a local minimum at $\phi=\phi_f$ and a global minimum at $\phi=\phi_t$. Note that a generalization of this problem to more than one field is rather straightforward by replacing the $\phi$ field with $\Phi=(\phi_1,\phi_2,\ldots,\phi_N)$. The path integral expression for the partition function is given by 
\begin{equation}\label{partition}
Z=\int \mathcal{D} \phi e^{-S_E[\phi]}. 
\end{equation}
In a saddle point approximation, the order parameter $\phi$ is to expand around its classical equation of motion (EOM) as $\phi=\bar{\phi}+\varphi$. Since $\bar{\phi}$ is a solution of EOM, it will minimise the Euclidean action and dominate the path integral. Thus, in the vicinity of the saddle point, the Euclidean can be written in term of a Taylor series up to the second order as
\begin{equation}\label{Euaction2}
S_E[\phi[\simeq S_E[\bar{\phi}]+\frac{1}{2} S_E''[\bar{\phi}]\varphi^2+\ldots,
\end{equation}   
where we note that the first-order variation is absent as the equations of motion. The second term in the above equation is a Gaussian integral, which can be evaluate exactly. After performing the functional Gaussian integration in the partition function in Eq.(\ref{partition}), one gets the following expression
\begin{equation}\label{Euaction3}
Z\sim e^{-S_E[\bar{\phi}]} \left( \mathrm{Det} S_E''[\bar{\phi}]  \right)^{-1/2}.
\end{equation}   
Then our next ultimate task is to find the solution of the EOM and substitute back into the partition function to obtain the free energy of the system through the standard formula $\mathcal{F}=-T \ln Z$. The nucleation rate $\Gamma$ could in turn be determined exactly from this Euclidean observable, the free energy, as $\Gamma= -2 \mathrm{Im}(\mathcal{F})$.

To compute the classical Euclidean field equation of motion, we extremize the Euclidean action
\begin{equation}\label{eom1}
\frac{\delta S_E[\phi]}{\delta \phi}\bigg|_{\phi=\bar{\phi}}=0.
\end{equation} 
Then the classical equation of motion becomes
\begin{equation}\label{eom2}
\left(\frac{\partial^2}{\partial \tau^2}+\nabla^2   \right)\bar{\phi}=V'(\bar{\phi}),
\end{equation}  
and solutions of this EOM with minimum energy are expected to be spherically symmetric \cite{Coleman:1977th}, then the critical field configuration of the field $\phi$ has $O(4)$ symmetry. However, for sufficiently high temperature, as argued by Linde \cite{Linde:1981zj}, the problem becomes approximately three dimensional by performing the integration over the Euclidean time coordinate. Accordingly, the Euclidean action is now rewritten as
\begin{equation}\label{Euaction4}
S_E[\bar{\phi}]\equiv\frac{S_3 [\bar{\phi}]}{T},
\end{equation}    
where $S_3 [\bar{\phi}]$ is the rescaled action of a three-dimensional field theory. In this case, the saddle point will be given by the $O(3)$ symmetric solution of  
\begin{equation}\label{eom3}
\frac{d^2 \bar{\phi}(r)}{dr^2}+\frac{2}{r}\frac{d \bar{\phi}(r)}{dr}=V'(\bar{\phi}).
\end{equation}
This is a second-order partial differential equation and has many solutions that extremize the Euclidean action. In the semi-classical approximation for false vacuum decay, not all saddle points are taken into account, what we are really interested in is a nontrivial saddle-point solution which interpolated between the false vacuum $\phi_f$ and some field value on the other side of the barrier. The field configuration $\phi_b$, often called bounce or instanton, is associated with the nucleation bubble embedded in the homogeneous false vacuum. The interior of the bubble consists of the true vacuum, whereas, outside the bubble, the field $\bar{\phi}$ should arrive at its false vacuum. Therefore, for the bounce, it is reasonable to employ the boundary condition of the false vacuum as $\lim\limits_{r \to \infty }\bar{\phi}(r)=\phi_f$. Since the EOM in Eq.(\ref{eom3}) is a second-order ordinary differential equation, another boundary condition $\frac{d\bar{\phi} (r)}{d r}|_{r=0}=0$ should be imposed on this EOM by the requirement of finite energy at the origin.    

Once we obtain the bounce $\phi_b$, the explicit expression for the three-dimensional Euclidean action $S_3$ is given by 
\begin{equation}\label{Eu3action}
S_3=4 \pi \int dr r^2 \left[ \frac{1}{2} \left(\frac{d \phi_b}{dr}  \right)^2+V(\phi_b) \right],
\end{equation}
and the surface tension, an one-dimensional energy of the bounce, is evaluated accordingly as
\begin{equation}\label{surfacet}
\Sigma=\int dr \left[ \frac{1}{2} \left(\frac{d \phi_b}{dr}  \right)^2+V(\phi_b) \right].
\end{equation}
In the practical calculations, It is worth noting that if the false vacuum has a non-zero potential energy, an additional term $-V(\phi_f)$ should be included in the $S_3$ action and the surface tension $\Sigma$. Finally, based on a saddle-point approximation around the bounce solution, the false vacuum will decay to the true vacuum with the rate per unit time per unit volume given at finite temperature by \cite{Linde:1981zj,Coleman:1977py,Callan:1977pt} 
\begin{equation}\label{nuclrate}
\Gamma= \frac{\omega_-}{2\pi}\left( \frac{S_3}{2\pi T} \right)^{3/2}\left[ \frac{\mathrm{Det}'(-\nabla^2+V''(\phi_b))}{\mathrm{Det}(-\nabla^2+V''(\phi_f))} \right]^{-1/2} e^{\left(-\frac{S_3}{T}  \right)}.
\end{equation}   
Here $\mathrm{Det}'$ indicates that the zero eigenvalues related to the translation symmetry of the bubble are ignored, $\omega_-$ is the eigenvalue of the negative mode and $\phi_b$ denotes the bounce solution of the EOM. Traditionally, the evaluation and computation of the prefactor in Eq.(\ref{nuclrate}) is a nontrivial work, but for the most typical situations, the Euclidean action presented in the exponent is quite large, consequently the exponential is very small. Hence the bounce solution and its action $S_3$ are the most important ingredient for the nucleation rate, the prefactor can be roughly estimated by dimensional analysis and then sketchily indicated as $T^4$ or $T^4_c$ for simplicity \cite{Csernai:1992tj,Scavenius:2000bb}. 

\subsection{The minima of the effective potential}

To give an intuitive description of thermal phase transition in particle and nuclear physics, the classical potential $V(\phi)$ should be generalized to the effective potential $V_{\mathrm{eff}}(\phi; T, \mu)$ in order to incorporated thermal and quantum corrections. Then for a given temperature and chemical potential, the effective potential $V_{\mathrm{eff}}$ is a function of the order parameter $\phi$. The expected value or the condensate for the order parameter is decided by minimizing the effective potential with the order parameter field, $\partial V_{\mathrm{eff}}/\partial \phi=0$. If there have many extrema, we need to calculate the second derivative to make sure it is a minimum point not a maximum point in the full parameter space. This is the standard procedure to search the minima of the effective potential for a single order parameter field.  

However, if there are many order parameters ($N>1$) in the effective potential, the above method is to be extended to a more general but complicated procedure as follows. Firstly, the gap equations will be obtained by minimizing $V_{\mathrm{eff}}(\phi_1,\phi_2,\ldots,\phi_N)$ with respect to the order parameters $\phi_i$ ($i=1,2,\ldots,N$),
\begin{equation}\label{gap1}
  \frac{\partial V_{\mathrm{eff}}}{\partial \phi_1}=  \frac{\partial V_{\mathrm{eff}}}{\partial \phi_2}=\ldots=0.
\end{equation}
The solutions of this equation give the points where the effective potential has its extrema in the full order parameter space. Secondly, the $N\times N$ Hessian matrix, 
\begin{equation}\label{gap2}
\mathcal{H}_{i,j}=\frac{\partial^2 V_{\mathrm{eff}}}{\partial \phi_i \partial \phi_j},
\end{equation}
has to be evaluated at the points where the gap equation (\ref{gap1}) is satisfied. At last, as long as we get the Hessian matrix, the points corresponding to the minima of the effective potential can be selected by checking the signs of the eigenvalues of the Hessian matrix. In the case of all positive values, such a point is a minimum, otherwise, it is either saddle point or a maximum. In a practical calculation, this method works very well while the effective potential is a "good'' potential with a large gradient around the minimum point, but the method becomes less efficient or even fails to give out the minima when the effective potential becomes very flat, such as the effective potential nearby the CEP. Besides the above standard method to explore the minima of the effective potential \cite{Mintz:2012mz}, for the physical interest in the present study, we propose an alternative method to find the minima of the effective potential based on a geometric approach, especially the local minima existed in the potential.

For the first step, for a potential with more than one order parameters $V(\Phi)$, supposed we already know one of the minimum of the potential, namely $\bar{\Phi}=(\bar{\phi}_1,\bar{\phi}_2,\ldots,\bar{\phi}_N)$. Here, we suggest to chose the point at the global minimum of the potential because it is easy to locate by scanning the potential directly in the order parameter space. In what follows, we denote these values as the expectation values. The standard procedure is to describe as follows: (1) the $N$-dimensional potential can be extremely simplified by treating the $\phi_1$ field as a variable while fixing other fields on their expectation values $\bar{\phi}_i$ ($i=2,\ldots,N$). Now, the $N$-dimensional problem has been changed to an one-dimensional problem, and it is easy to dig up other local minima in the $\phi_1$ field direction, i.e. $\phi_1=\phi_1^{\alpha}, \phi_1^{\beta}, \ldots$. (2) the $\phi_2$ field is considered as a variable whereas other fields maintain their expectation values all the time, then the local minima of the effective potential $V(\phi_2)$ can also be found as the case for a single order parameter field. These minima of the potential $V(\phi_2)$ are expressed as $\phi_2=\phi_2^{\alpha}, \phi_2^{\beta}, \ldots$. (3) by using the same algorithm, we can find all minima in the direction of the $\phi_3$ field as $\phi_3=\phi_3^{\alpha}, \phi_3^{\beta}, \ldots$, and so on. (4) after obtaining the minima for the potential in every direction, the landscape of the potential can be roughly reconstructed by combining the potentials in different direction all together. In particular, if there is only one minimum in some directions, those parameters can be discard and the numbers of the dimension is to be reduced accordingly. 

For the second step, the primary points obtained through the above algorithm in the first step are not necessary the true local minima of the potential. To trace a true local  minimum of the potential, we have to increase the dimension of the parameter space or the number of the order parameter one by one. (1) starting from the initial point at $V=V(\phi_1^{\alpha(0)}, \phi_2, \bar{\phi}_3, , \ldots)$, by taking the $\phi_2$ field as a variable, we can find the minima of the potential when other fields are fixed at $\phi_1^{\alpha}=\phi_1^{\alpha(0)}$ and $\phi_i=\bar{\phi}_i$ for $i=3,\ldots,N$. Let us assume one of the minima is $\phi_2^{\alpha(0)}$. (2) we treat the $\phi_1$ as a variable when fixing other fields at $\phi_2=\phi_2^{\alpha(0)}$ and $\phi_i=\bar{\phi}_i$ for $i=3,\ldots,N$. If one of the minima of the potential at $\phi_1=\phi_1^{\alpha(1)}$ is equal to the original point at $\phi_1=\phi_1^{\alpha(0)}$, we can conclude that the point ($\phi_1^{\alpha(0)}$, $\phi_2^{\alpha(0)}$) at the ($\phi_1, \phi_2$) plane is the true local minimum of the potential in the $2$-dimensional parameter space. Otherwise, we need to repeat the first stage again by setting $\phi_1=\phi_1^{\alpha(1)}$ in the potential in order to gain $\phi_2^{\alpha(1)}$. After several iterations, we can luckily arrive at a right point when the value of the $\phi_1$ satisfies $\phi_1^{\alpha(k)}=\phi_1^{\alpha(k+1)}$, then the point ($\phi_1^{\alpha(k)}$, $\phi_2^{\alpha(k)}$) is the exact local minimum what we are looking for. Otherwise, after several iterations, if the value of the $\phi_1^{\alpha(k+1)}$ is far away from its previous value at $\phi_1=\phi_1^{\alpha(k)}$ with the increase of the number of iteration $k$, the starting point ($\phi_1^{\alpha(0)}$, $\phi_2^{\alpha(0)}$) is a saddle point, it should be discarded accordingly. (3) in the last stage, once we get the minimum point in the $\phi_1$ and $\phi_2$ plane, akin to the first and second stages, we can extract the local minimum of the potential in a $3$-dimensional order parameter space by employing the field $\phi_3$ as a variable. Ultimately, one by one, step by step, we can gain all minima of the potential in the full order parameter space.     

\subsection{The tunneling between two minima}

Generally, for a given potential, it is a tough job to find all minima in the $N$-dimensional order parameter space if          
$N\geq 2$. As discussed in above subsection, we can use two different algorithms to do calculations. For the first algorithm, it is straightforward and easy to apply. The advantage of this algorithm is that we can use it to find all minima of the potential very efficiently and quickly. But for some problems, since we don't know the geometry of the potential, even though we have exact positions of two minima, it does not help to find the instanton connected with these two minima. For example, there is a two-dimensional potential in Fig.2 in Ref. \cite{Masoumi:2016wot} that does not have any instanton solution between these two local minima. On the other side, for the second algorithm developed by us, it is complicated or even awkward to be applied, due to the fact that we have to find a local minimum of the potential step by step. Nevertheless, the reward for such a sophisticated procedure is that we can change a multi-dimensional problem to a one-dimensional problem, which is usually easy to solve. More specifically, since the tunneling occurs in two nearby minima, in the practical calculation, we actually do not need to know all minima of the potential. These two minima which are close to each other and can be connected through the instanton are more important and crucial. Therefore, to study the problem of the false vacuum decay, we prefer to adopt the second algorithm to find two minima lived nearby.     

To demonstrate the algorithm developed in the above section, we firstly take an interesting two-dimensional potential which has many minima as a typical example. The potential which is formerly proposed in Ref. \cite{Masoumi:2016wot} has the form
\begin{equation}\label{example1}
  V(\phi_1,\phi_2)=\sin(\phi_1-\phi_2)+\frac{1}{2}\cos(\phi_1+\phi_2)+\cos3(\phi_1+\phi_2)+2\cos3(\phi_1-\frac{\phi_2}{2}).
\end{equation}  
There are many minima in this potential, but we need not to know all of them in the parameter space. What we are interested are those they are close to each other. Assume we already know one of the minima at the point $p_1=(2.39338,2.82768)$, our next task is to find out the other minimum close to this one. Following the second algorithm developed in above, by taking $\phi_1$ as a variable, the potential $V(\phi_1)$ has three local minima at $(\phi_1^{\alpha}=0.376583, \phi_1^{\beta}=2.39338, \phi_1^{\gamma}=4.56047)$ when fixed the second order parameter at $\phi_2=2.82768$. This implies that there have three local minima in the direction of the $\phi_1$ field. In the meantime, if we fix $\phi_1=2.39338$, the potential $V(\phi_2)$ displays others three minima in the direction of the $\phi_2$ field. So that there probably exist four local minima around the point $p_1$. It is worth noting that if we start from the initial point $p_1$, we can only find the four minimum points around this point, however, if we run over the minimum points in the potential one by one, the missing counting problem can be resolved and all missing minimal points can be recovered accordingly. Next, as mentioned in above, the initial point $(\phi_1^{\alpha(0)}=0.376583,\phi_2^{\alpha(0)}=2.39338)$ could not be a true minimum in the potential. We need to take the $\phi_2$ as a variable to search the local minima of the potential when fixing the $\phi_1$ field at ${\phi_1^{\alpha(0)}=0.376583}$. After a simple calculation, we can find one of the minimum of the potential is at $\phi_2^{\alpha(1)}=2.74049$. Then by fixing the $\phi_2$ field at $\phi_2^{\alpha(1)}=2.74049$, we can obtain a minimum point at $\phi_1^{\alpha(1)}=0.376074$. Through the iteration method, we can get $\phi_2^{\alpha(2)}=2.74048$ and $\phi_1^{\alpha(2)}=0.376074$. In this time, since we have $\phi_1^{\alpha(2)}=\phi_1^{\alpha(1)}$, we can conclude the point at $p_2=(\phi_1^{\alpha(2)},\phi_2^{\alpha(2)})=(0.376074,2.74048)$ is a true local minimum of the potential in the $\phi_1-\phi_2$ plane. Accordingly, the tunneling rate between these two minima $p_1$ and $p_2$ is described by the system of coupled ordinary differential equations
\begin{eqnarray}
  \frac{d^2 \phi_1}{dr^2}+\frac{2}{r}\frac{d \phi_1}{dr} &=& \frac{\partial V(\phi_1,\phi_2)}{\partial \phi_1}, \\
 \frac{d^2 \phi_2}{dr^2}+\frac{2}{r}\frac{d \phi_2}{dr} &=& \frac{\partial V(\phi_1,\phi_2)}{\partial \phi_2}, 
\end{eqnarray}
with the boundary conditions at $\lim\limits_{r \to \infty }\phi(r)=p_1$ and $\frac{d\phi (r)}{d r}|_{r=0}=0$ because the point $p_2$ has a lower potential than $p_1$ and is regarded as a true vacuum. 

For a second example, we turn back to our present study, the effective potential $\Omega_{\mathrm{MF}}(T,\mu)$ in Eq.(\ref{potential}) now has three order parameters. Traditionally, it is hard to give out an intuitive landscape picture of the potential. Hence, it is necessary to simplify a multi-dimensional problem to an one-dimensional problem. According to the second algorithm, for a given temperature and chemical potential, we firstly find the global minimum of the effective potential by either solving the gap equations or scanning the effective potential directly. The exact values of the order parameters in the global minimum of the potential are called as the expectation values for the $\sigma$, $\Phi$ and $\Phi^*$ fields. But the big challenge is to find the possible local minima nearby this global one. Following the second algorithm, we can treat the $\sigma$ field as a variable, but fixed other parameters $\Phi$ and $\Phi^*$ at their expectation values, and the results have been depicted in Fig.\ref{Fig03}. On the contrary, the order parameter $\Phi$ can also be taken as a variable, but setting the $\sigma$ and $\Phi^*$ fields on their expectation values instead as shown in Fig.\ref{Fig02}. From these figures, it can be found that the local minima of the effective potential only exists in the direction of the $\sigma$ field and there is no minimum in both $\Phi$ and $\Phi^*$ directions. Therefore, we can dissociate one of the equation of motion for the $\sigma$ field from a system of three coupled ordinary differential equations very neatly. Since the effective potentials for the $\Phi$ and $\Phi^*$ fields always exhibit an ``$U$" type, the equations of motion have two trivial solutions which are the expectation values. 

Consequently, the bounce is then a solution to the equation of motion for the $\sigma$ field when taking the $\Phi$ and $\Phi^*$ fields on their expectation values
\begin{equation}\label{eom_s}
 \frac{d^2 \sigma}{dr^2}+\frac{2}{r}\frac{d \sigma}{dr} = \frac{\partial \Omega_{\mathrm{MF}}}{\partial \sigma},
\end{equation}       
with the boundary conditions $\lim\limits_{r \to \infty }\sigma(r)=\sigma_F$ and $\frac{d\sigma (r)}{d r}|_{r=0}=0$. Here, $\sigma_F$ is the false vacuum or the local minimum of the effective potential. It is worth noting that our analyses are different from the previous works in Refs. \cite{Mintz:2012mz,Stiele:2016cfs}, where the first-order hadron quark phase transition is also considered in the three-flavor PQM model. In their works, the jumps of the order parameters for the strange quark and the Polyakov-loops fields were treated as active variables when the temperature is close to a first-order coexistence line. Based on the basic picture of the thin-wall approximation, they have parameterized the bounce solution through the interpolation of two values of the order parameter corresponded to two minima of the effective potential for four order parameters, i.g $\sigma_x$ for $u$ $d$ quarks, $\sigma_y$ for $s$ quark, $\Phi$ and $\Phi*$. However, according to our above analyses, the jumps of the order parameters for the Polyakov-loops fields are fake signal to be taken as a first-order phase transition. These sudden leaps are induced by the disconnection of the chiral order parameter $\sigma$ and do not support by the effective potential. Moreover, our analyses are also in agreement with the results in Refs. \cite{Mao:2009aq,Gupta:2011ez,Schaefer:2011ex,Tiwari:2013pg}, where the deconfinement phase transition and the phase transition for the $s$ quark are crossover. Therefore, we would like to treat the Polyakov-loops fields as a background fields rather than the active variables during the chiral phase transition.

\section{Results and discussion}

In this section, a bounce is to be gained by numerically solving the equation of motion in Eq.(\ref{eom_s}) with the proper boundary conditions, $\sigma\rightarrow \sigma_{F}$ as $r\rightarrow \infty$ and $\sigma'(0)=0$. In this case, a three-dimensional problem has been simplified to an one-dimensional problem when we constrain the order parameter space in the $\sigma$ field direction, thus the numerical package \cite{Masoumi:2016wot} and the discussions presented in previous studies in the QM model \cite{Wang:2023omt} can be applied straightforwardly.  

For a typical first-order phase transition, the effective potential $\Omega_{\mathrm{MF}}$ in the $\sigma$ field direction displays three distinct extrema. One is the potential barrier, two of them are the local minima representing the quark and hadron phases, respectively. For convenience, the quark phase is denoted as $\sigma_l$, whereas the hadron phase is defined as $\sigma_h$. As $T=T_{\chi}^c$, these two minima are degenerate, however, with the increase of the temperature further, the quark phase at $\sigma=\sigma_l$ becomes the absolute minimum of the effective potential and chiral symmetry is approximately restored for $T>T_{\chi}^c$. So that the quarks lose most of their constituent masses and become almost massless in this phase. Simultaneously, the other minimum at $\sigma=\sigma_h$ is the local minimum, and it is taken as a false vacuum. On the other side, as the temperature is to drop down from its critical temperature $T_{\chi}^c$, the hadron phase at $\sigma=\sigma_h$ is the absolute minimum of the effective potential and should be considered as the true vacuum, whereas the quark phase becomes a metastable state and is treated as a false vacuum. Since the false vacuum and true vacuum will get flipped when the temperature goes across to its critical value $T_{\chi}^c$, $\sigma_F$ in the boundary conditions should be set as $\sigma_F=\sigma_l$ for $T>T_{\chi}^c$ and  $\sigma_F=\sigma_h$ for $T<T_{\chi}^c$. Under these boundary conditions, in what follows, we will divide our studies into two categories: a weak first-order phase transition and a strong one for sake of convenience.     

\subsection{A weak first-order phase transition}

In the region of the weak first-order phase transition when the chemical potential is less than $\mu_c\simeq 309$ MeV, as the temperature is lowered from its critical value at $T_{\chi}^c$, the local minimum at $\sigma=\sigma_l$ approaches the intervening maximum, then two extrema meet and form an inflection point at the spinodal temperature $T_{\mathrm{sp}}\equiv T_{\mathrm{c1}}$. Thus, there only exists one minimum in the effective potential for $T<T_{\mathrm{c1}}$. Similar results can also be applied to the case in another direction of the temperature. When the temperature is larger than the critical one, $T_{\chi}^c$, with the increase of the temperature, the local minimum at $\sigma=\sigma_h$ will move up to the intervening maximum and eventually these two extrema will merge together at another spinodal temperature $T_{\mathrm{sp}}\equiv T_{\mathrm{c2}}$, so that only the global minimum remains for $T>T_{c2}$. Correspondingly, when the temperature is among two spinodal temperatures $T_{\mathrm{c1}}$ and $T_{\mathrm{c2}}$, there has a false vacuum, therefore there could exist a bounce solution to connecting the false vacuum and the true vacuum, otherwise, we have only a trivial solution  for the equation of motion, i.e. the expectation value of the order parameter.  

\begin{figure}[thbp]
\epsfxsize=9.0 cm \epsfysize=6.5cm
\epsfbox{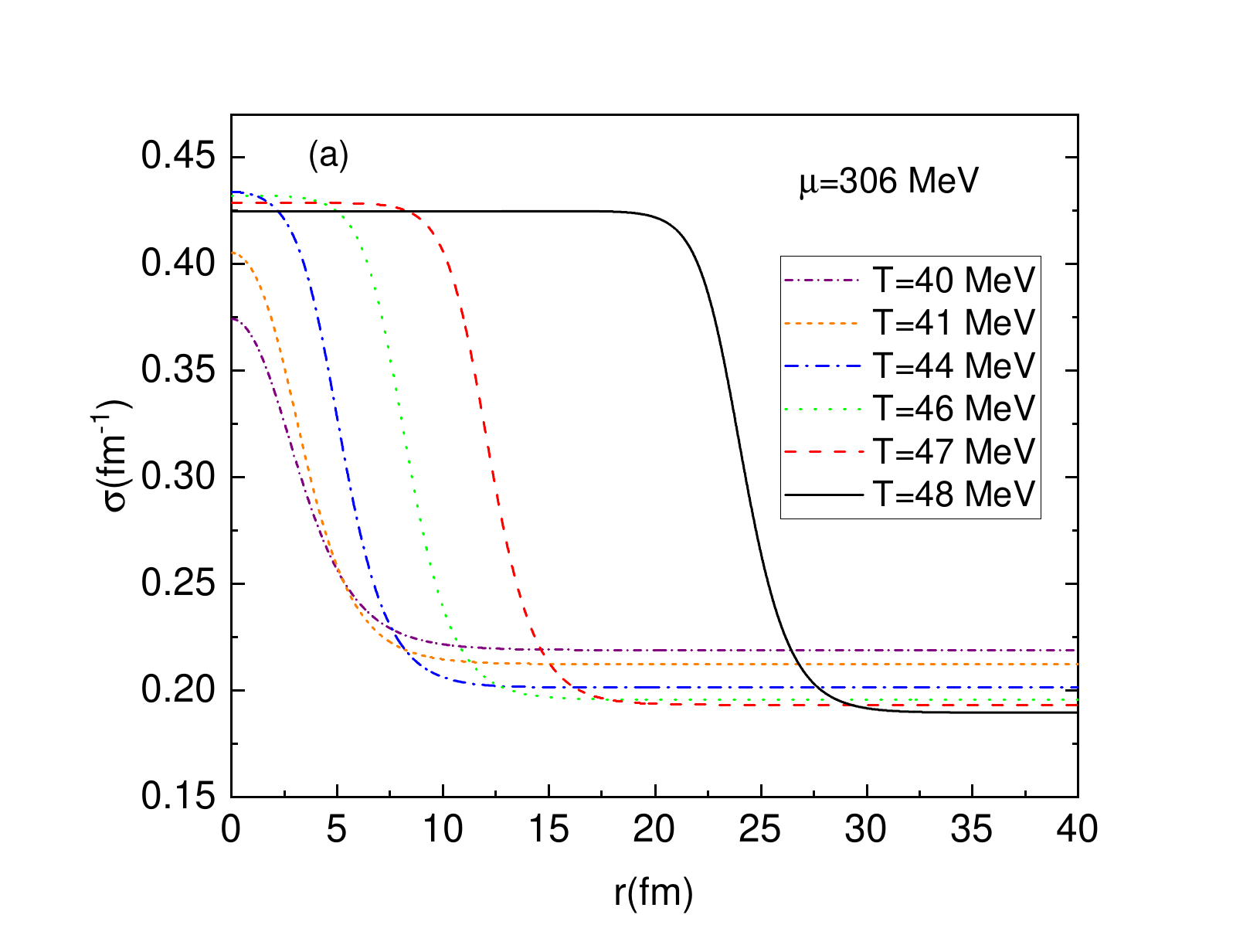}\hspace*{0.01cm} \epsfxsize=9.0 cm
\epsfysize=6.5cm \epsfbox{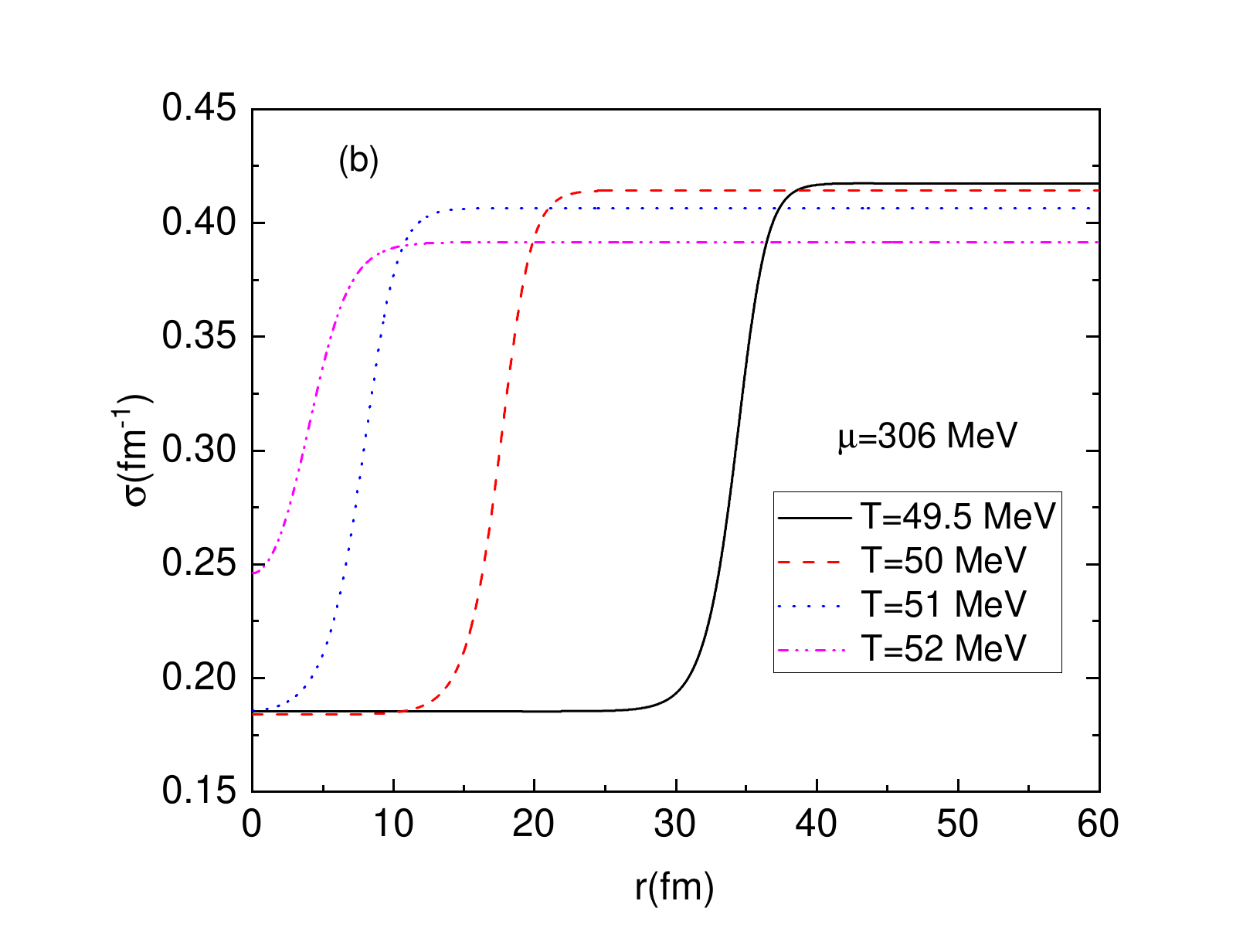}
 \caption{(Color online) (a) Bubble profiles for different temperatures when fixing the chemical potential at $\mu=306$ MeV for $T<T_{\chi}^c$. From left to right, the curves correspond to $T=40$, $41$, $44$, $46$, $47$ and $48$ MeV. (b) Bubble profiles for different temperatures when fixing the chemical potential at $\mu=306$ MeV for $T>T_{\chi}^c$. From right to left, the curves correspond to $T=49.5$, $50$, $51$ and $52$ MeV.}
\label{Fig05}
\end{figure}

For $\mu=306$ MeV, the exact numerical solutions of the equation of motion (\ref{eom_s}) are plotted in the left panel of Fig.\ref{Fig05} as $T=40$, $41$, $44$, $46$, $47$ and $48$ MeV when the temperature is below the coexistence line, $T<T_{\chi}^c\simeq 48.9$ MeV. Here, the specific boundary conditions are selected as $\sigma\rightarrow \sigma_l$ as $r\rightarrow \infty$ and $d \sigma(0)/dr=0$. Form this figure, as the temperature is very close to the critical value at $T=T_{\chi}^c$, the bounce solutions show an obvious ``core" structure with $\sigma\simeq \sigma_h$ inside the bubble which is separated by a relatively thin wall from the outside false vacuum at $\sigma\simeq \sigma_l$. While the temperature approaches to another limit for the existence of the bounce solution, $T_{c1}\simeq 28$ MeV, the bubble profiles usually become a ``coreless" structure due to the fact that the radii of the bubbles are comparable to the thickness of the walls, thereby the thin-wall approximation can not be applied. On the other side of the coexistence line as $T>T_c$. the bubble profiles are obtained by solving the equation of motion (\ref{eom_s}) with the boundary conditions $\sigma\rightarrow \sigma_h$ as $r\rightarrow \infty$ and $d \sigma(0)/dr=0$. The results are plotted in the right panel of Fig.\ref{Fig05} when taking the temperatures at $T=49.5$, $50$, $51$ and $52$ MeV for $\mu=306$ MeV. For the temperature between critical value at $T_{\chi}^c\simeq 48.9$ MeV and the up spinodal temperature at $T_{c2}\simeq 52.1$ MeV, the bubble profiles exhibit an uniform structure with the true vacuum inside separated by a relatively thin wall from the outside false vacuum, only if the temperature is very close to its critical one at $T=T_{\chi}^c$, the center of the bubble is to deviate from its true vacuum value at $\sigma=\sigma_l$ largely since the thickness of the bubble wall is of the same order as the radius.        

\begin{figure}[thbp]
\epsfxsize=9.0 cm \epsfysize=6.5cm
\epsfbox{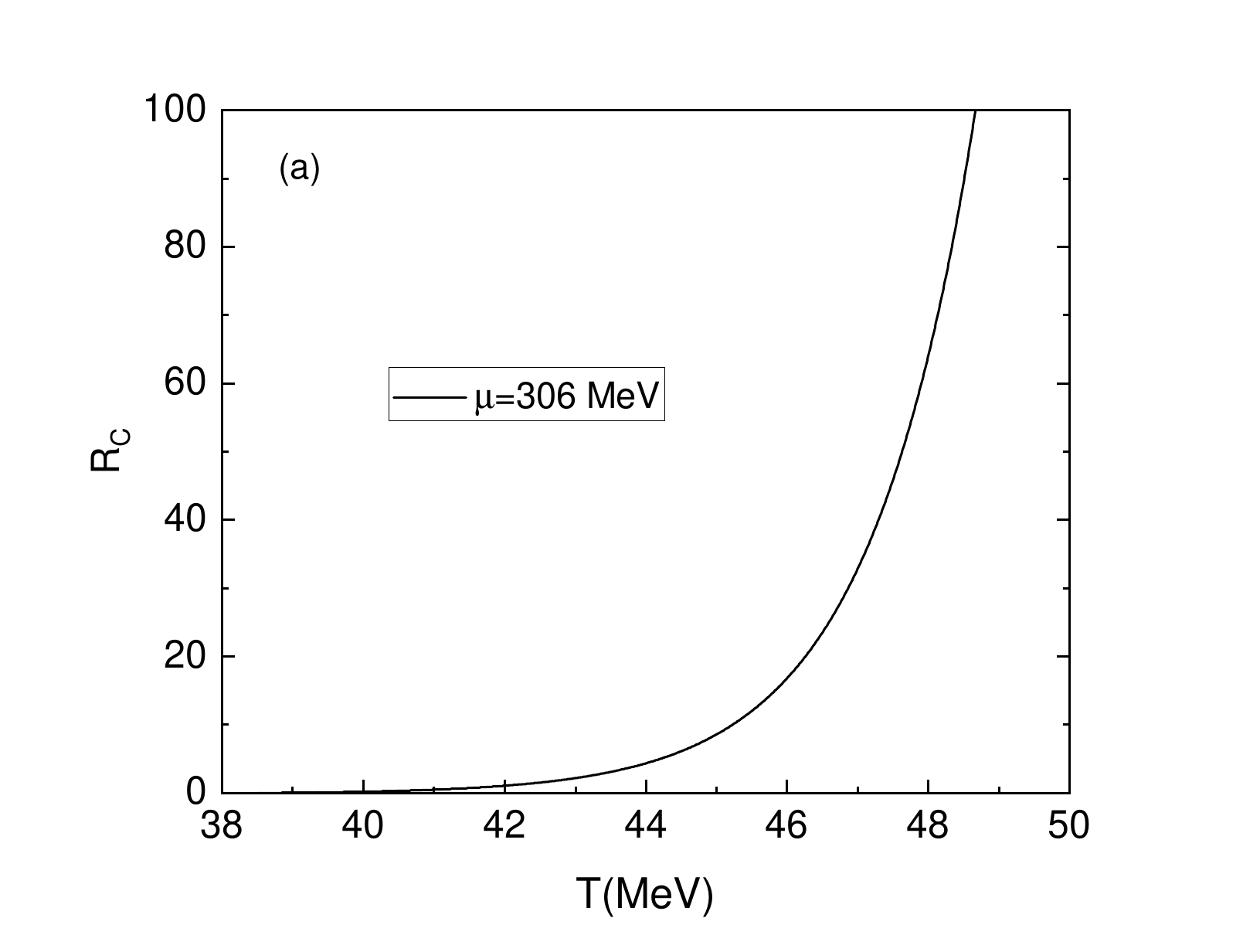}\hspace*{0.01cm} \epsfxsize=9.0 cm
\epsfysize=6.5cm \epsfbox{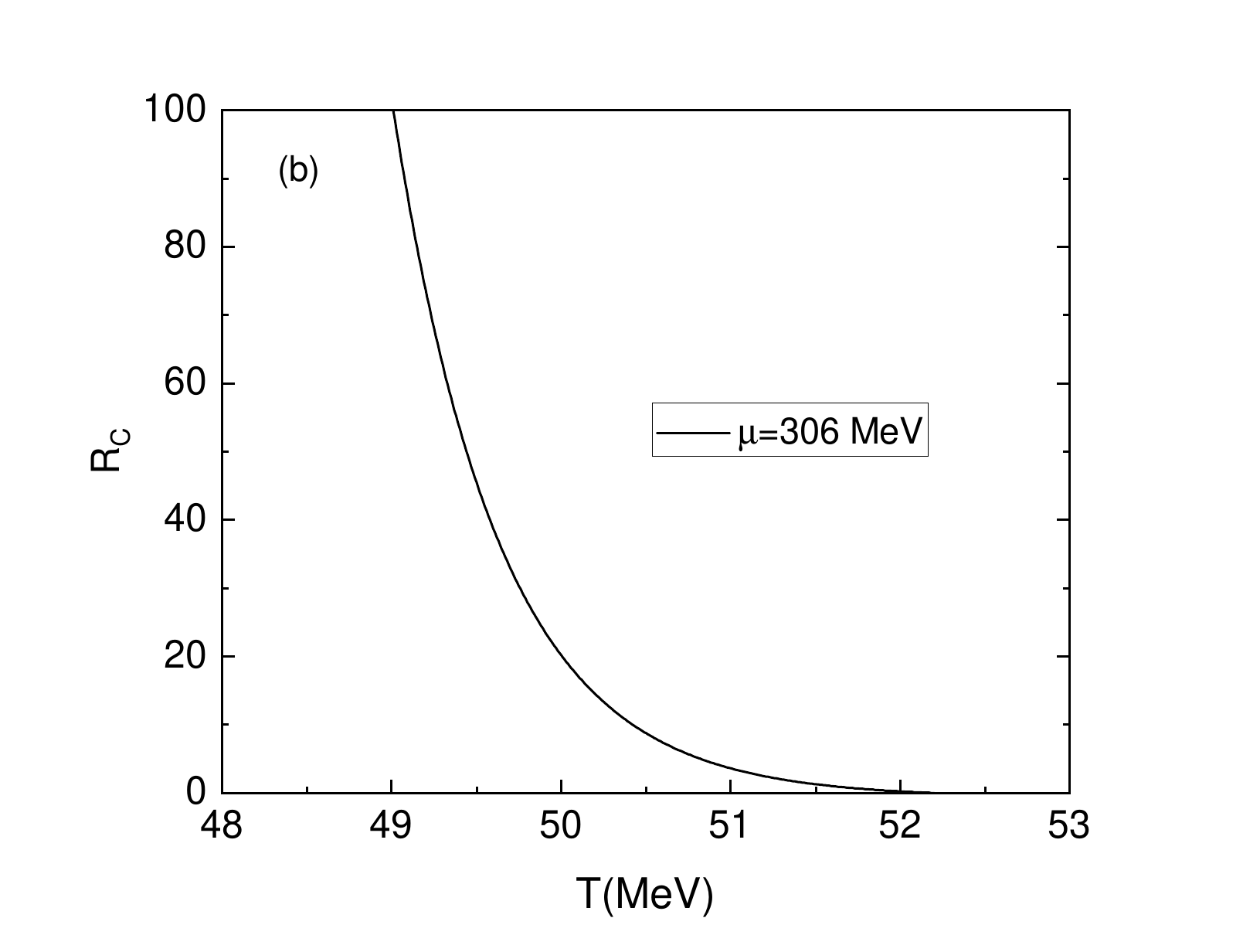}
 \caption{(Color online) (a) The radius of the bounce as a function of temperature $T$ when $T<T_{\chi}^c$ for $\mu=306$ MeV. (b) The radius of the bounce as a function of temperature $T$ when $T>T_{\chi}^c$ for $\mu=306$ MeV.}
\label{Fig06}
\end{figure}

To give a more specific description of the size of the bubble, the typical radius of the bounce can be roughly estimate by the maximal value of the first derivative of the $\sigma(r)$ field, which gives $R_c\equiv\left | \sigma'(r) \right |$. It is worth pointing out that this kind of definition of $R_c$ is rather arbitrary and we should not confuse with the critical radius $\mathcal{R}_c$ defined in the thin-wall approximation. In the framework of the thin-wall approximation, when the radius is much larger than the thickness of the bubble wall, characterized by $\xi \simeq 1/m_{\sigma}$, the frictional force $2 \sigma'/r $ in the equation of motion (\ref{eom_s}) can be neglected and the three-dimensional Euclidean action $S_3$ is well approximated by the expression
\begin{equation}\label{s3_thinwall}
 S_3=4 \pi r^2 \Sigma-\frac{4}{3} \pi r^3 \varepsilon,
\end{equation}
where $\varepsilon=\Omega(\sigma_F)-\Omega(\sigma_T)$ is the energy difference between the false vacuum and the true vacuum. From this equation (\ref{s3_thinwall}), there is an energy contest among the energy consumption from the creation of an interface of the bubble and the energy enhancement from the phase conversion. If the bubble is small, the energy cost is higher than the energy gain, the bubble will shrink and disappear. On the contrary, a very large bubble usually reflects a larger energy increase than the energy cost. Therefore, there must exist a bubble with the critical size according to the energy competition. So that the critical radius of the bubble is defined by minimization of the action $S_3$ with respect to $r$ as $\mathcal{R}_c=2\Sigma/ \varepsilon$.  

In Fig.\ref{Fig06}, typical radii of the bounces as a function of temperature $T$ are obtained for $\mu=306$ MeV in the two metastable regions. From this figure, the typical radius goes to zero at the spinodal line for both cases when the temperature is above or below the coexistence line. The reason is because the false vacuum is to become unstable and the phase conversion occurs via the spinodal decomposition process rather than the bubble nucleation, hence there is no bounce solution anymore as $T\rightarrow T_{\mathrm{sp}}$. This is an obvious feature of the weak first-order phase transition and it seems this character does not appear in the thin-wall approximation, where the typical radius does not vanish but only becomes very small \cite{Scavenius:2000bb,Palhares:2010be}. However, when the temperature closes in the critical line at $T=T_{\chi}^c$, the exact numerical calculation and the thin-wall approximation show a similar behavior: the radius of the bounce rise up sharply and become divergent as $T\rightarrow T_{\chi}^c$. This is due to the fact that the radius of the bounce incline to expand very quickly as the temperature $T$ approaches to $T_{\chi}^c$, when the radius is large enough, the thin-wall approximation is applied, we can end up with the consequence, $R_c\sim \mathcal{R}_c$. From the definition of the critical radius $\mathcal{R}$, noting that $\varepsilon\rightarrow 0$ as  $T\rightarrow T_{\chi}^c$, the radius of the bounce should also become divergent when $T=T_{\chi}^c$. 

\begin{figure}[thbp]
\epsfxsize=9.0 cm \epsfysize=6.5cm
\epsfbox{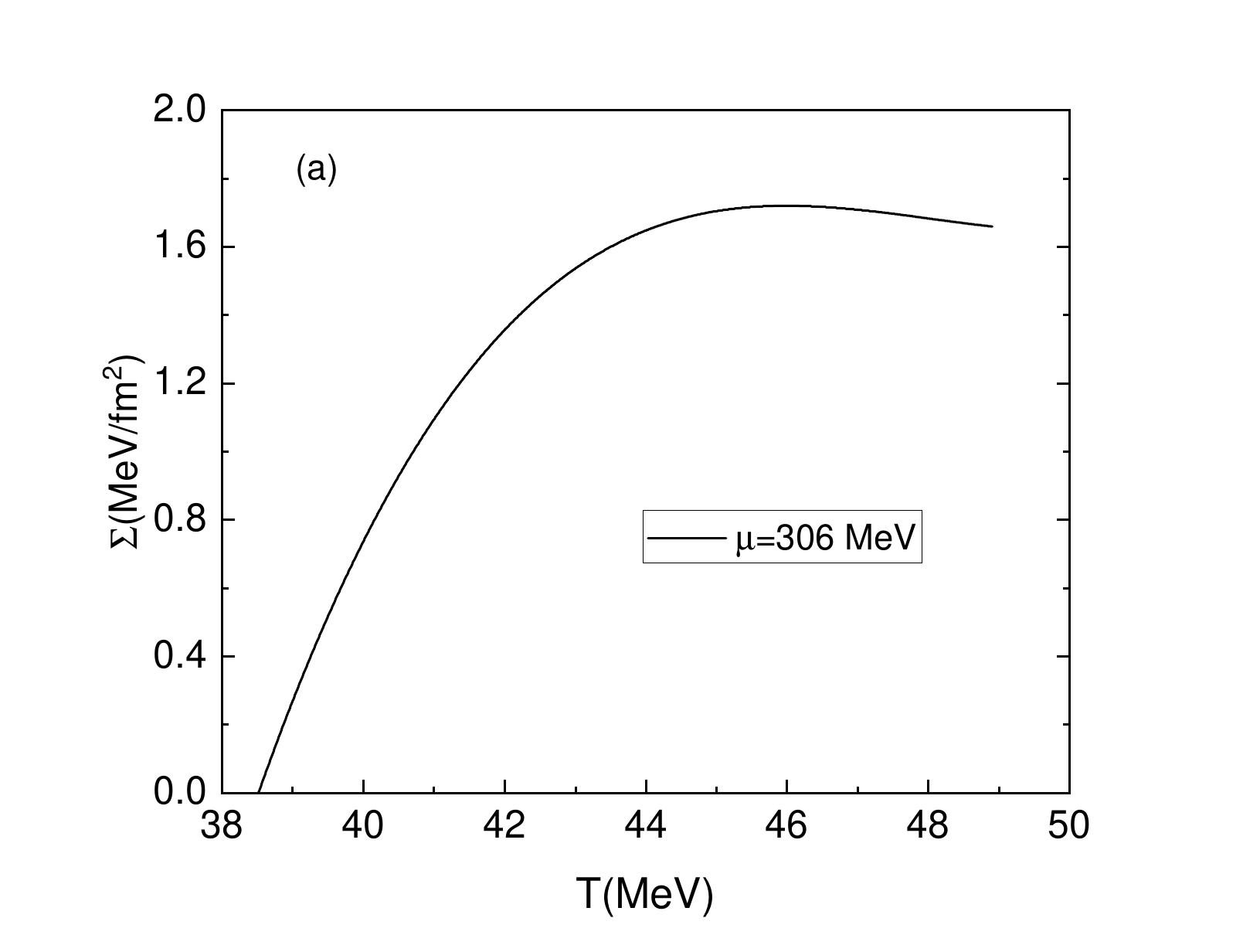}\hspace*{0.01cm} \epsfxsize=9.0 cm
\epsfysize=6.5cm \epsfbox{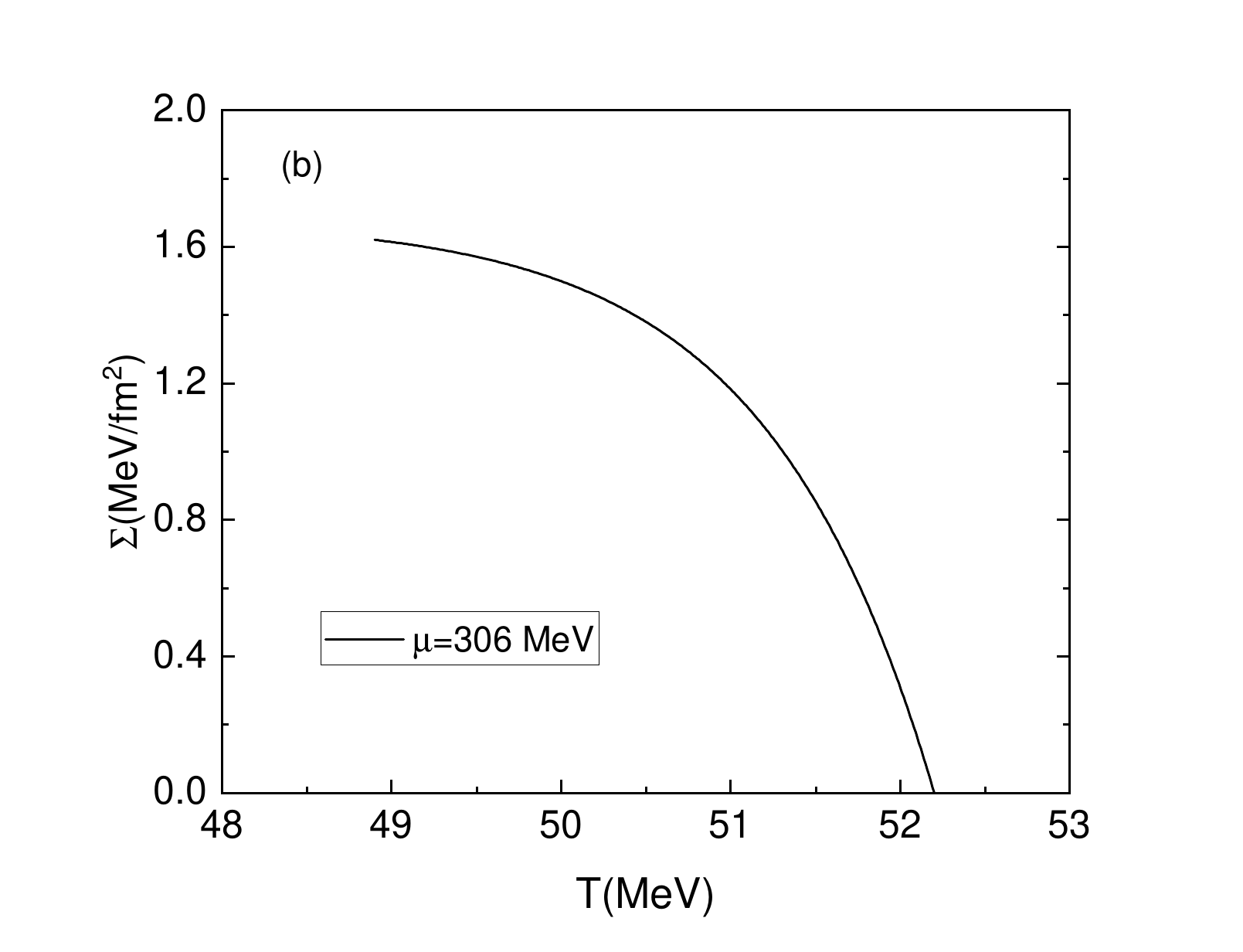}
 \caption{(Color online) (a) Surface tension as a function of temperature $T$ when $T< T_{\chi}^c$ for $\mu=306$ MeV. (b) Surface tension as a function of temperature $T$ when $T> T_{\chi}^c$ for $\mu=306$ MeV.}
\label{Fig07}
\end{figure}

The surface tension plays a central role in determining the process of the bubble nucleation because it represents the amount of energy per unit of area in the creation of an interface between the two vacua. Once the bounce solutions have been obtained, the surface tension of the nucleation bubble can be calculated by using the definition in the equation (\ref{surfacet}). In Fig.\ref{Fig07}, we show the surface tension $\sigma$ as a function of the temperature $T$ for $\mu=306$ MeV as the temperature is among the up and down spinodal lines. For both cases, as the temperature leads to its spinodal critical value $T_{\mathrm{sp}}$, the surface tension reduces to zero since there is no potential barrier and we can only have a trivial solution to the equation of motion at that moment. Oppositely, the surface tension $\Sigma$ shows a quite different behavior in the hadron phase by comparison with that of the quark phase. When the temperature is below the coexistence line, with the increase of the temperature, the surface tension $\Sigma$ starts to grow from zero dramatically and then reaches a maximal value at a certain temperature, after that it will inflect and bend downwards. Such a nontrivial behavior of the surface tension was also reported for a first-order hadron-quark phase transition in the QM model \cite{Wang:2023omt,Palhares:2010be} and the Friedberg-Lee model \cite{Zhou:2020bzk}, where the exact numerical method has been applied. But the non-monotonic feature of the surface tension did not appear in the studies based on the thin-wall approximation \cite{Mao:2019aps,Bessa:2008nw,Mintz:2012mz}, so that the inflection point of the surface tension is usually taken as a hint for the breakdown of the thin-wall approximation. On the contrary, when $T>T_{\chi}^c$, the surface tension $\Sigma$ demonstrates a trivial character with a simply monotonic function. With the decrease of the temperature, it will grow up almost linearly from zero to a largest value as $T\rightarrow T_{\chi}^c$ given that there is a biggest potential barrier and a smallest free energy difference between two vacua at this point.  

\begin{figure}[thbp]
\epsfxsize=9.0 cm \epsfysize=6.5cm
\epsfbox{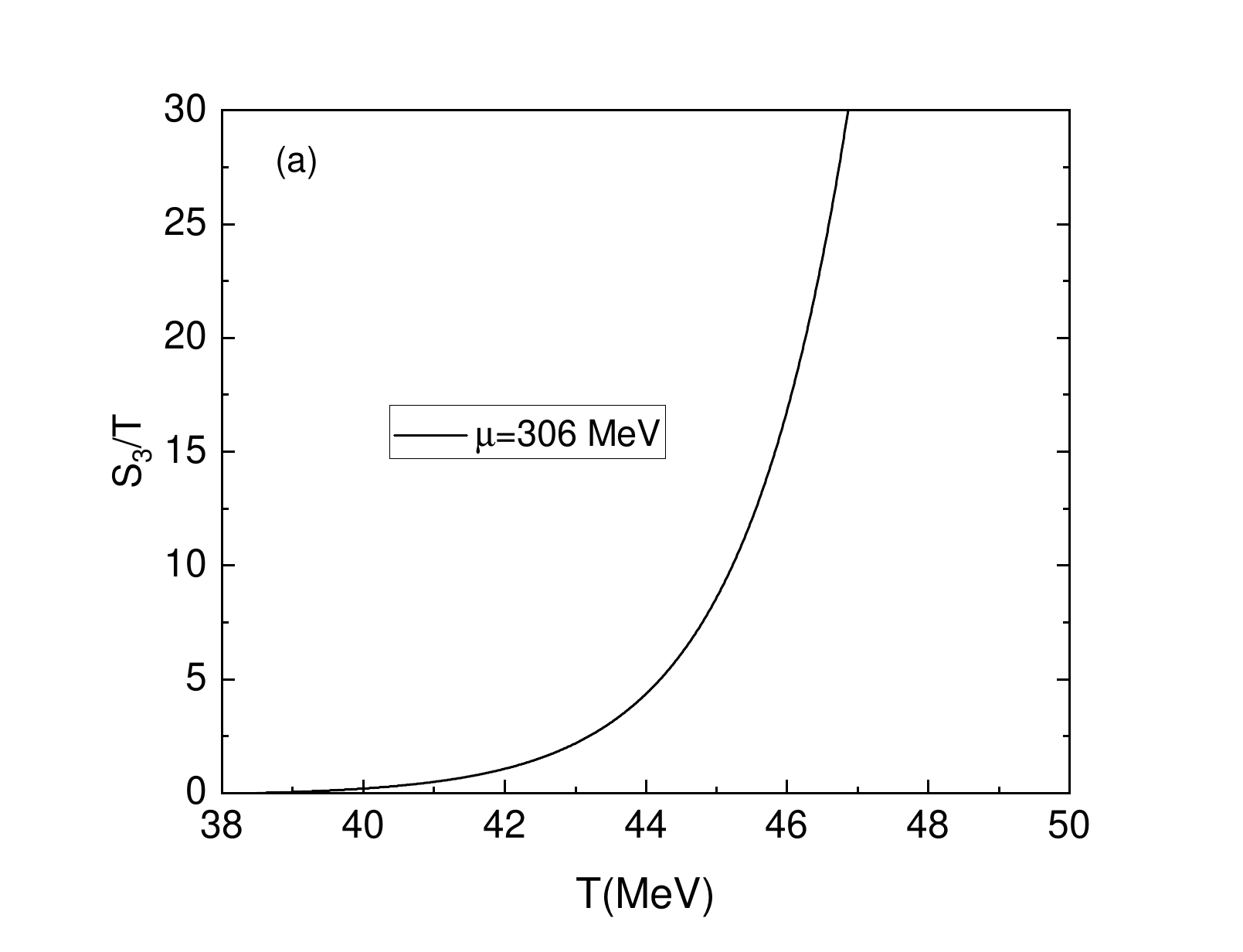}\hspace*{0.01cm} \epsfxsize=9.0 cm
\epsfysize=6.5cm \epsfbox{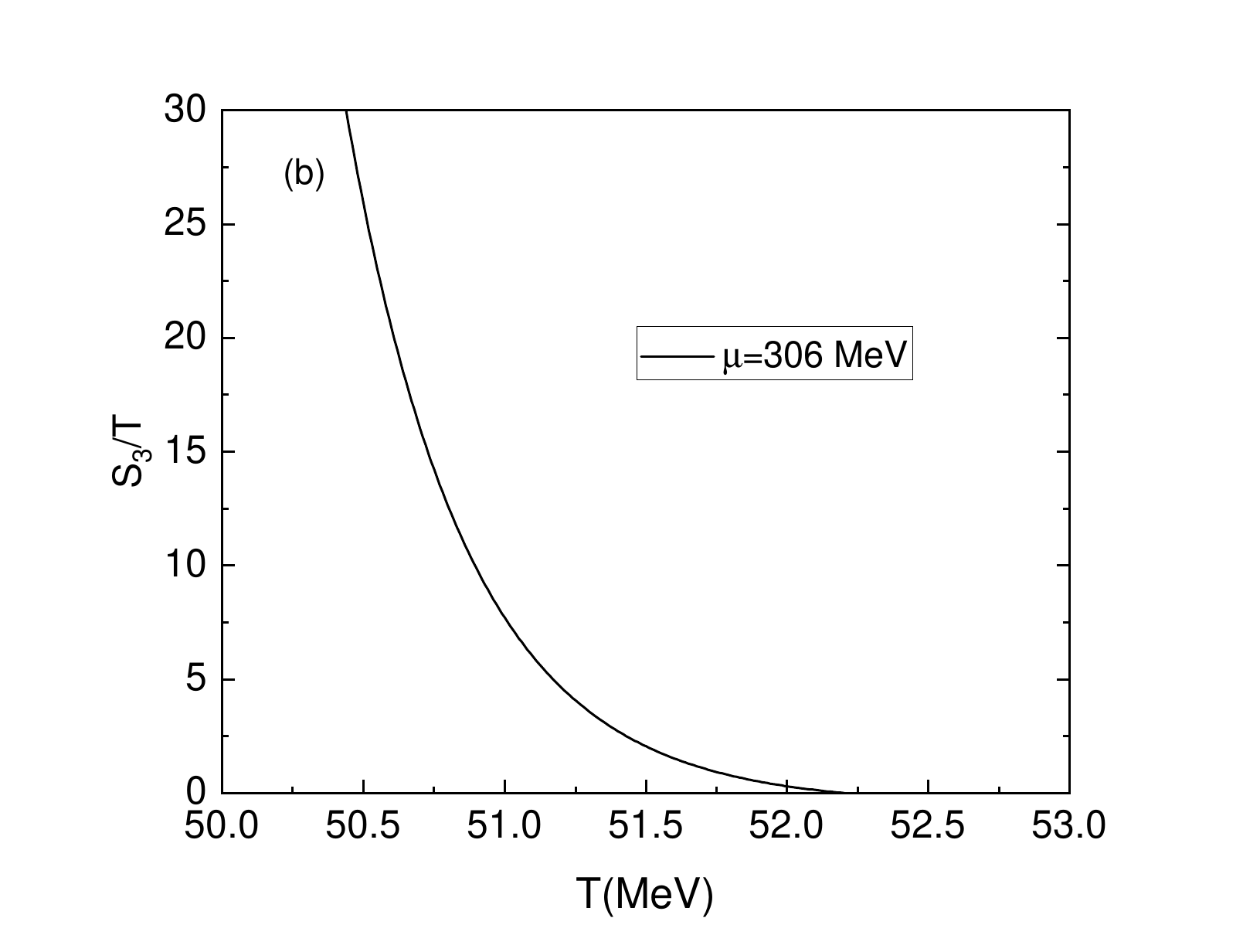}
 \caption{(Color online) (a) The saddle-point action evaluated on the bounce as a function of temperature $T$ when $T< T_{\chi}^c$ for $\mu=306$ MeV. (b) The saddle-point action evaluated on the bounce as a function of temperature $T$ when $T> T_{\chi}^c$ for $\mu=306$ MeV.}
\label{Fig08}
\end{figure}

To estimate the nucleation rate for homogeneous nucleation theory, the largest contribution to the result is the action $S_3$ evaluated at a saddle-point because it has the exponential form in the equation (\ref{nuclrate}). While the prefactor in the nucleation rate that comes from fluctuations around the saddle-point solution is of negligible numerical contribution in comparison with the term $e^{-S_3/T}$ and it can then be estimated based on dimensional analysis. Especially, when the saddle point action $S_3/T$ is much larger than the unit one. In order to display the saddle-point action due to the appearance of the bounce, the $S_3/T$ exponent as a function of temperature $T$ for $\mu=306$ MeV is plotted in Fig.\ref{Fig08} when the temperature is between the up and down spinodal critical temperatures. From this figure, the action will begin at the zero value as $T=T_{\mathrm{sp}}$, then it will rise up very quickly and become divergent as $T\rightarrow T_{\chi}^c$. Therefore, in the most area of the two metastable regions between the spinodal lines, the action $S_3/T$ is larger than the unit one and the decay of the false vacuum is still exponentially suppressed by the saddle-point action. Unless the temperature is very close to the spinodal critical value $T_{\mathrm{sp}}$, the action $S_3/T$ is to go acrose below to the unit one. This indicates that the hadron phase can survive in the region above the critical temperature line until the temperature is close to the up spinodal line or the quark phase can also live in the region below the critical temperature down to the another spinodal line. Thus, we believe it is better to use the spinodal line to separate the phase boundary of the quark and hadron phase rather than the critical coexistence line for the first-order phase transition.            

\subsection{A strong first-order phase transition}

\begin{figure}[thbp]
\epsfxsize=9.0 cm \epsfysize=6.5cm
\epsfbox{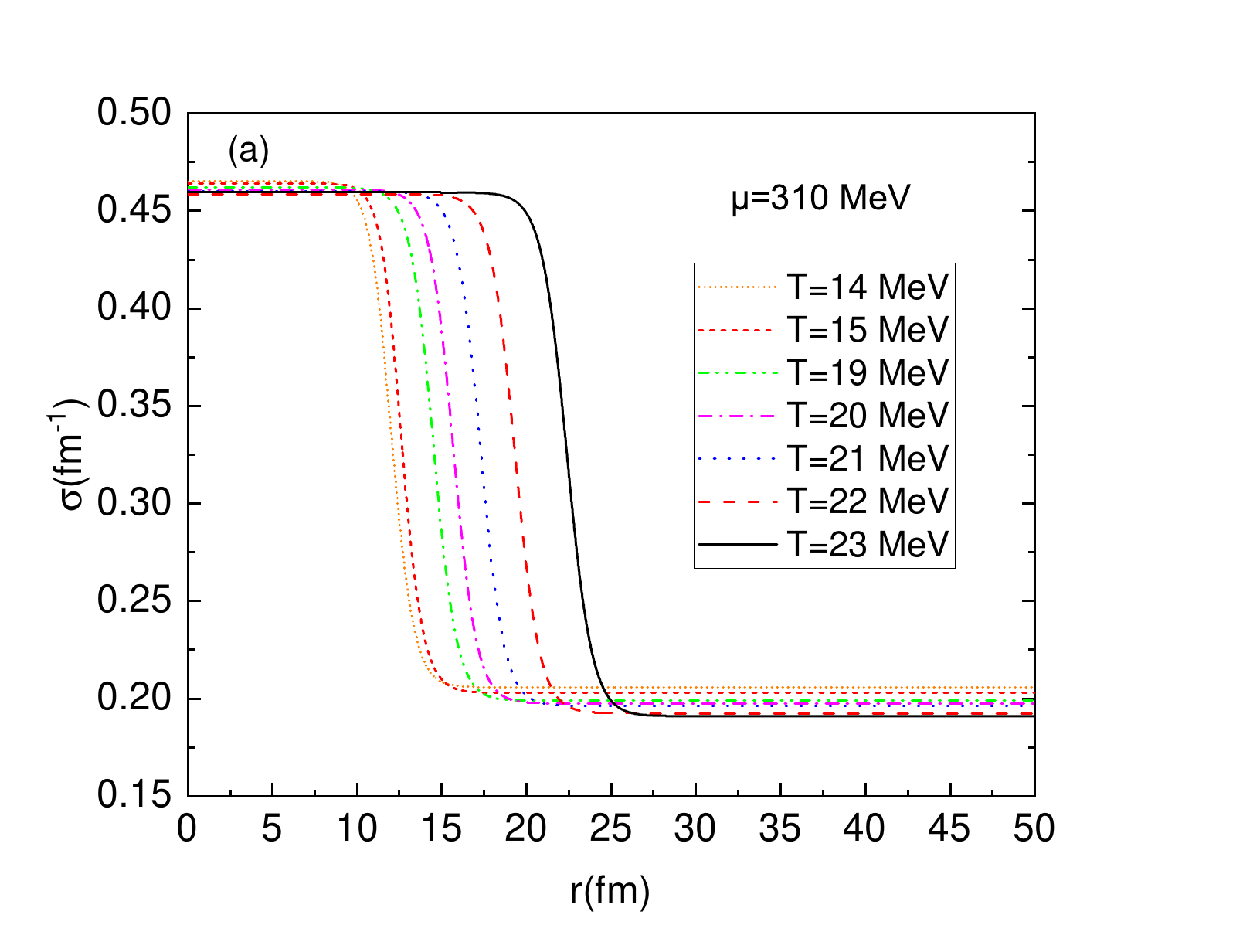}\hspace*{0.01cm} \epsfxsize=9.0 cm
\epsfysize=6.5cm \epsfbox{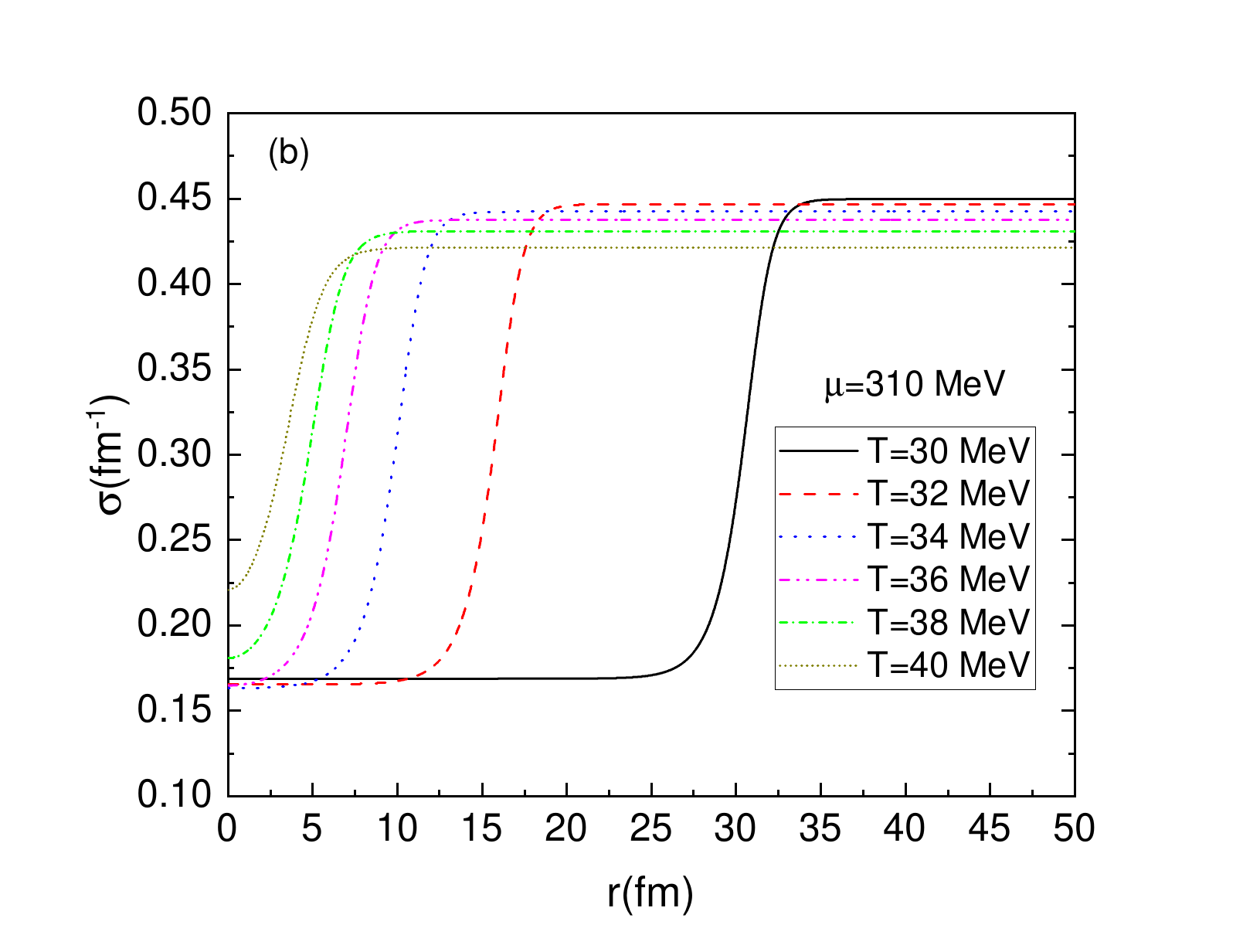}
 \caption{(Color online) (a) Bubble profiles for different temperatures when fixing the chemical potential at $\mu=310$ MeV for $T<T_{\chi}^c$. From left to right, the curves correspond to $T=14$, $15$, $19$, $20$, $21$, $22$ and $23$ MeV. (b) Bubble profiles for different temperatures when fixing the chemical potential at $\mu=310$ MeV for $T>T_{\chi}^c$. From right to left, the curves correspond to $T=30$, $32$, $34$, $36$, $38$ and $40$ MeV.}
\label{Fig09}
\end{figure}

As discussed in the above section, when the chemical potential $\mu$ is larger than $\mu_c\simeq 309$ MeV, there always exists a potential barrier even when the temperature goes to zero for $T<T_{\chi}^c\simeq 28.8$ MeV. Normally, we denote a phase conversion induced by an effective potential with a zero-temperature potential barrier as a strong first-order phase transition. Taking $\mu=310$ MeV as a typical example of a strong first-order hadron quark phase transition, we have numerically solved the equation of motion (\ref{eom_s}) with the proper boundary conditions and the results have been depicted in Fig.\ref{Fig09}. From the picture, as the temperature is close to the critical temperature $T_{\chi}^c$, the bounce solution usually exhibits a ``core" structure with the true vacuum inside and the false vacuum outside and the two vacua has been separated by a thin wall. This implicates that the thin-wall approximation is applicable for the temperature nearby the critical temperature. On the contrary, when the system is far away from the critical coexistence line, the bounce solution will lose its core structure due to the fact that the ``frictional' term in the equation of motion (\ref{eom_s}) becomes more and more important and can not be discarded as the decrease of the radius $r$. Moreover, since there is no spinodal line for the strong first-order phase transition, we can inevitably have the bounce solution which is the continues saddle-point field profile connected two vacua when $T<T_{\chi}^c$. Whereas on the other side of the critical coexistence line as $T>T_{\chi}^c$, with the increase of temperature, the potential barrier begin to decline and finally disappear at $T_{\mathrm{sp}}\simeq 41.4$ MeV. Hence there has only one minimum in the effective potential, thereby we have no bounce solution anymore as $T\geq T_{\mathrm{sp}}$.  

\begin{figure}[thbp]
\epsfxsize=9.0 cm \epsfysize=6.5cm
\epsfbox{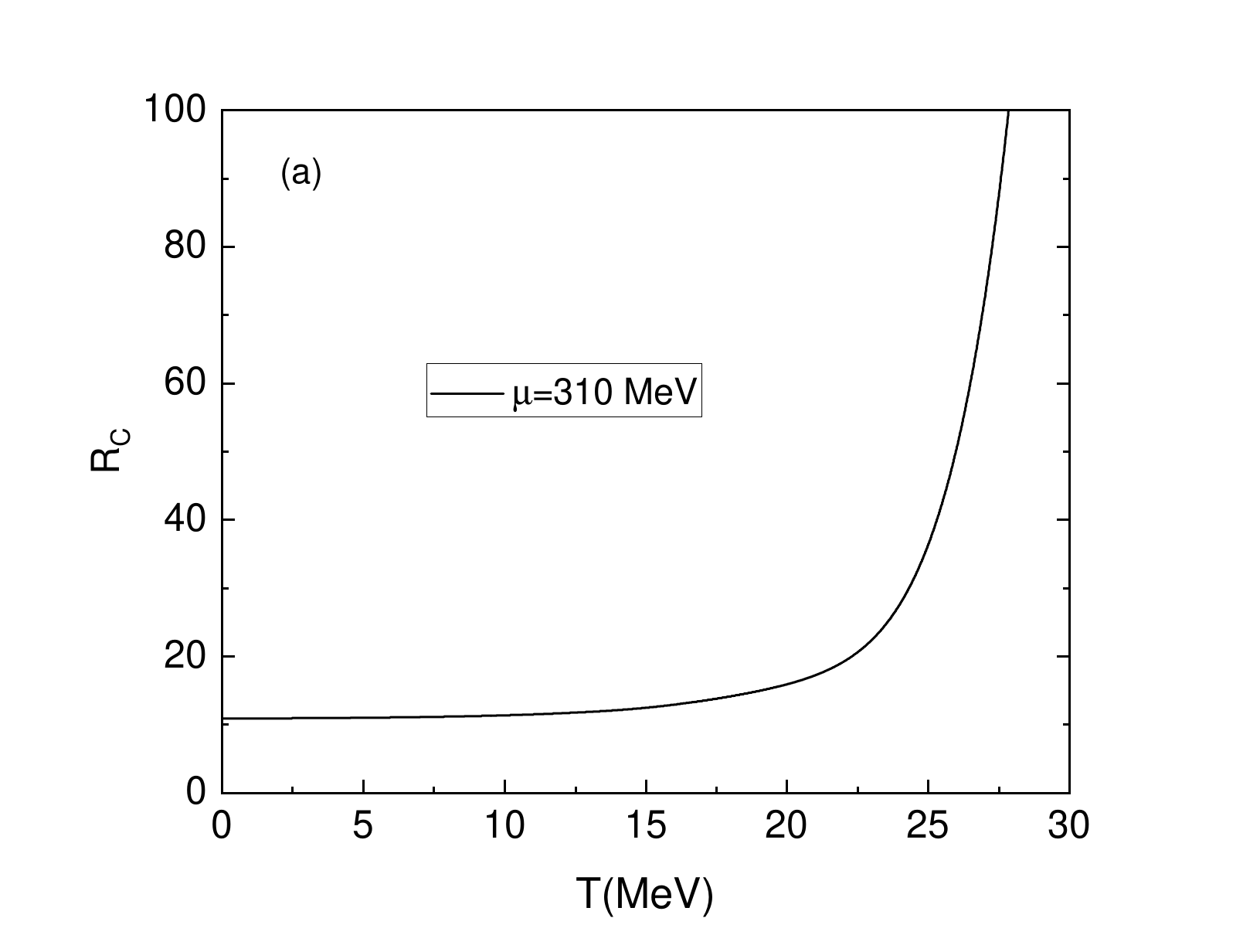}\hspace*{0.01cm} \epsfxsize=9.0 cm
\epsfysize=6.5cm \epsfbox{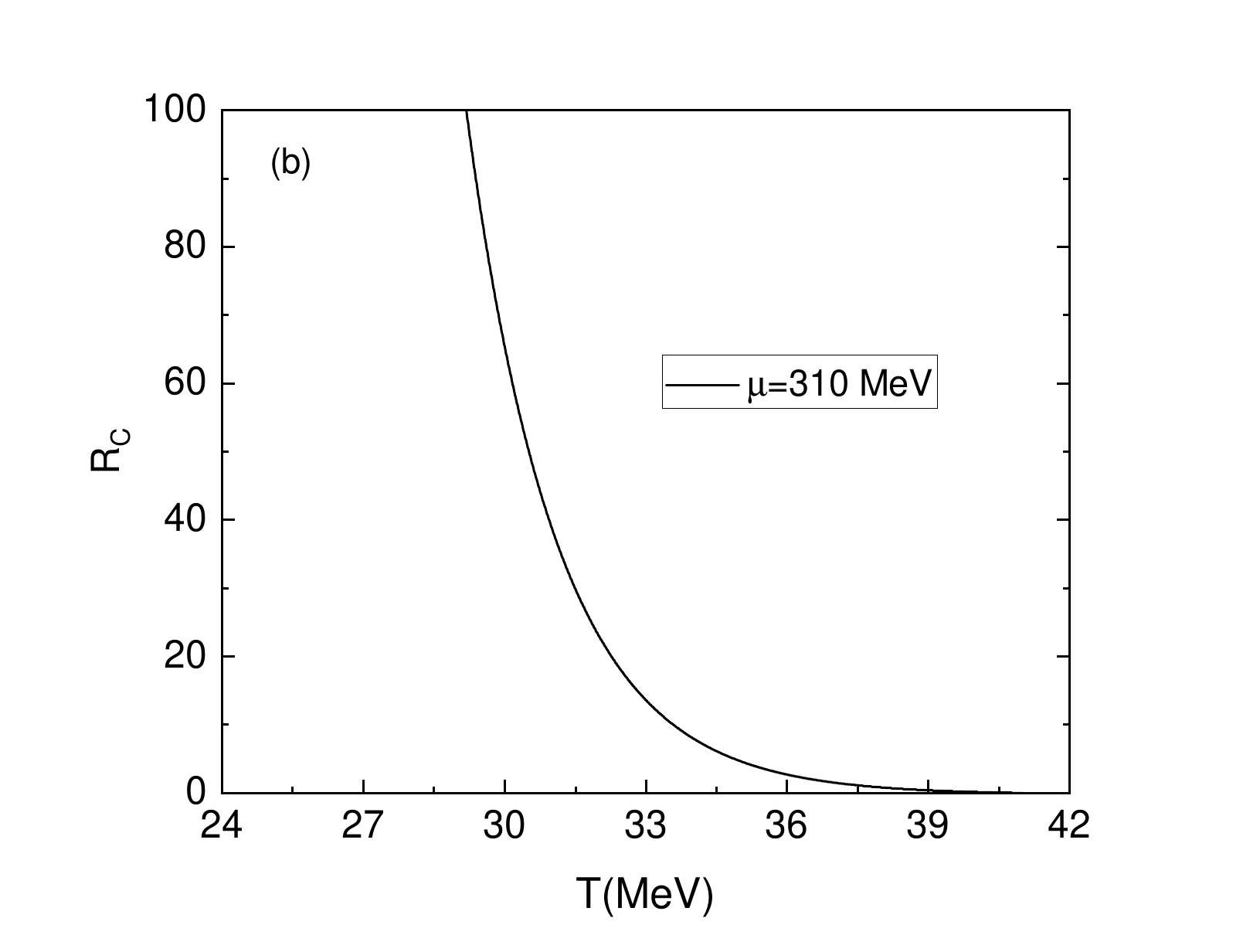}
 \caption{(Color online) (a) The radius of the bounce as a function of temperature $T$ when $T<T_{\chi}^c$ for $\mu=310$ MeV. (b) The radius of the bounce as a function of temperature $T$ when $T>T_{\chi}^c$ for $\mu=310$ MeV.}
\label{Fig10}
\end{figure}

In order to reveal the size of the bounce evolving with the temperature for a strong first-order phase transition, we have drawn the radius $R_c$ of the bounce as a function of the temperature $T$ both in the hadron phase and in the quark phase as depicted in Fig.\ref{Fig10}. In a traditional quark phase as $T>T_{\chi}^c$, while the temperature reduces from the spinodal line to a critical value, the radius of the bounce will increase sharply and become divergent at $T=T_{\chi}^c$ for the reason that two vacua get degenerate and the energy difference $\varepsilon$ is zero at that moment. On the other hand, when the system is located at the hadron phase, with the ascent of the temperature, the radius of the bounce climbs very slowly and steadily until the temperature is near a specific temperature somewhere at $T\simeq 20$ MeV. After that temperature, it will raise up dramatically and diverge at $T=T_{\chi}^c$. This is an apparent difference between a strong first-order phase transition and a weak first-order phase transition in the hadron quark phase conversion.      

\begin{figure}[thbp]
\epsfxsize=9.0 cm \epsfysize=6.5cm
\epsfbox{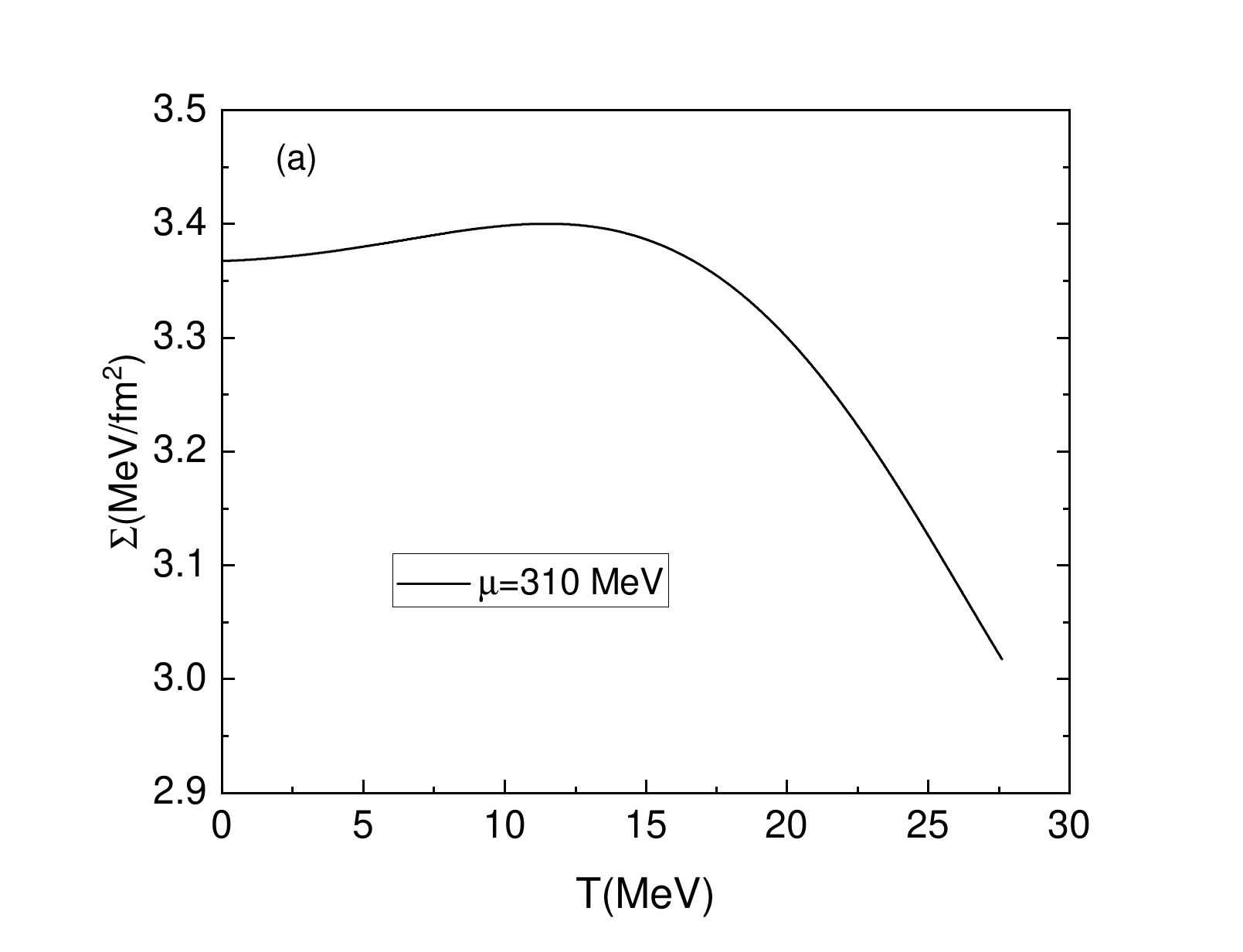}\hspace*{0.01cm} \epsfxsize=9.0 cm
\epsfysize=6.5cm \epsfbox{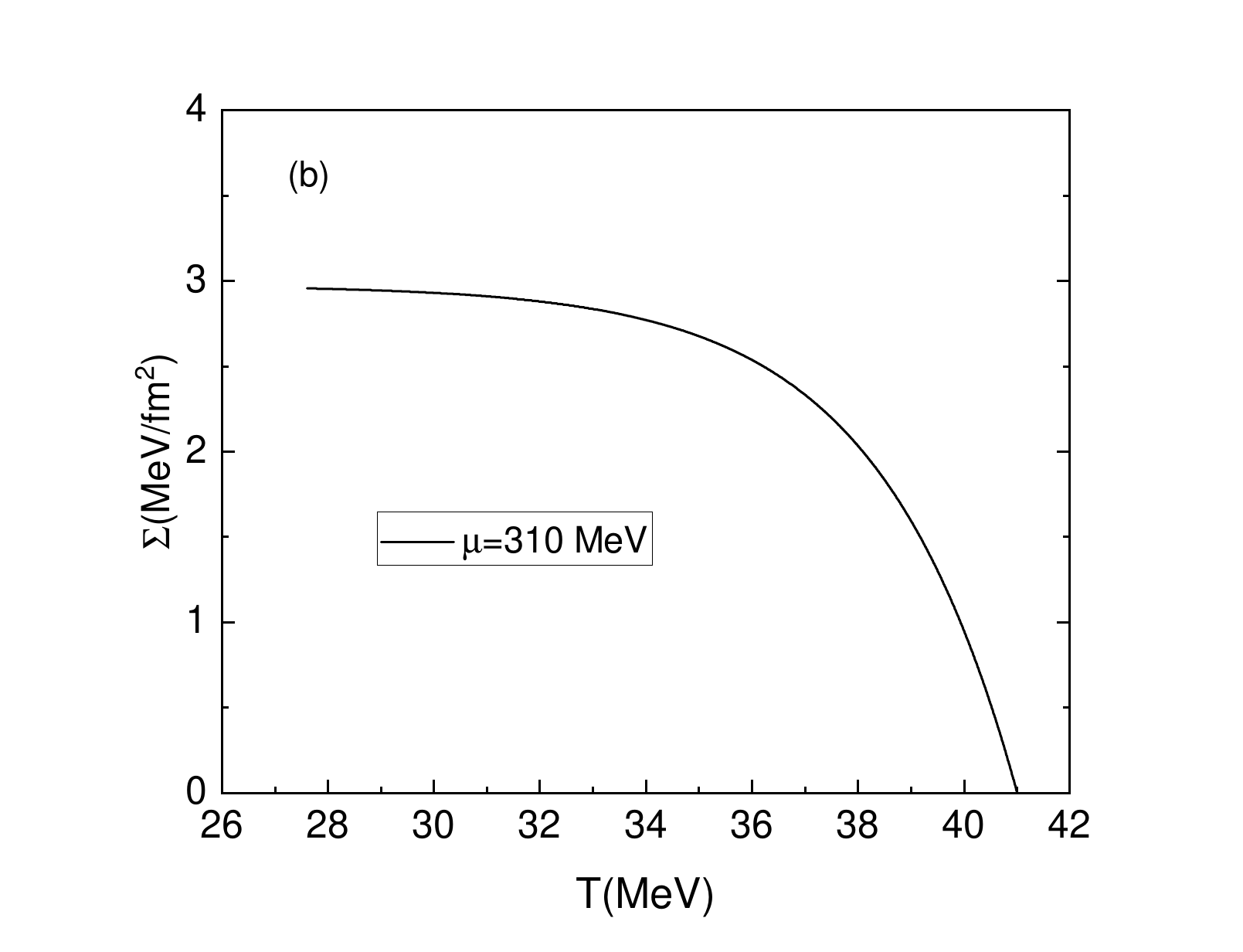}
 \caption{(Color online) (a) Surface tension as a function of temperature $T$ when $T< T_{\chi}^c$ for $\mu=310$ MeV. (b) Surface tension as a function of temperature $T$ when $T> T_{\chi}^c$ for $\mu=310$ MeV.}
\label{Fig11}
\end{figure}

Similar to the case of a weak first-order phase transition, the surface tension in the hadron phase also presents a non-monotonic behavior as shown in Fig.\ref{Fig11}.  With the increase of the temperature, it will grow up slightly and arrive at a maximum at a certain temperature. After that it starts decreasing rapidly to a minimum $\Sigma(T)\sim 3$ $\mathrm{MeV/fm^2}$ at $T=T_{\chi}^c$. As mentioned in the previous discussion, the turning point of the surface  in general suggests a limit for the application of the thin-wall approximation. However, from the Fig.\ref{Fig11}, there are two visible differences between a strong and a weak first-order phase transition. One is that the surface tension $\Sigma(T)$ as a function of the temperature does not goes to zero with the reduction of the temperature by reason of the existence of the bounce for a strong first-order phase transition as $T<T_{\chi}^c$. Meanwhile, according to Fig.\ref{Fig10}, since the radius of the bounce is almost a constant when the temperature is less than $18$ MeV, the surface tension carries an analogous property, it does not change too much with the reduction of the temperature when the temperature is smaller than $18$ MeV. Furthermore, the other obvious difference is the value of the surface tension at $\mu=310$ MeV is much larger than that of a weak first-order phase transition at $\mu=306$ MeV. As $T=T_{\chi}^c$, the former is about $3$ $\mathrm{MeV/fm^2}$, whereas the latter is only about half of that around $1.6$ $\mathrm{MeV/fm^2}$. Therefore, we can conclude that the surface tension will increase with the enhancement of the chemical potential due to the fact that a large chemical potential usually hints a large potential barrier. 

In a quark phase as $T>T_{\chi}^c$, the surface tension as a function of temperature $T$ has been demonstrated in the right panel of Fig.\ref{Fig11}. It is shown that the surface tension $\Sigma(T)$ monotonically goes down from a maximum $\sim3$ $\mathrm{MeV/fm^2}$ to zero when the temperature is ascent up to the spinodal line at $T=T_{\mathrm{sp}}$, where there is no potential barrier so that there is no bounce solution any longer. This trivial behavior of the surface tension is one of the obvious character of a weak first-order phase transition.

\begin{figure}[thbp]
\epsfxsize=9.0 cm \epsfysize=6.5cm
\epsfbox{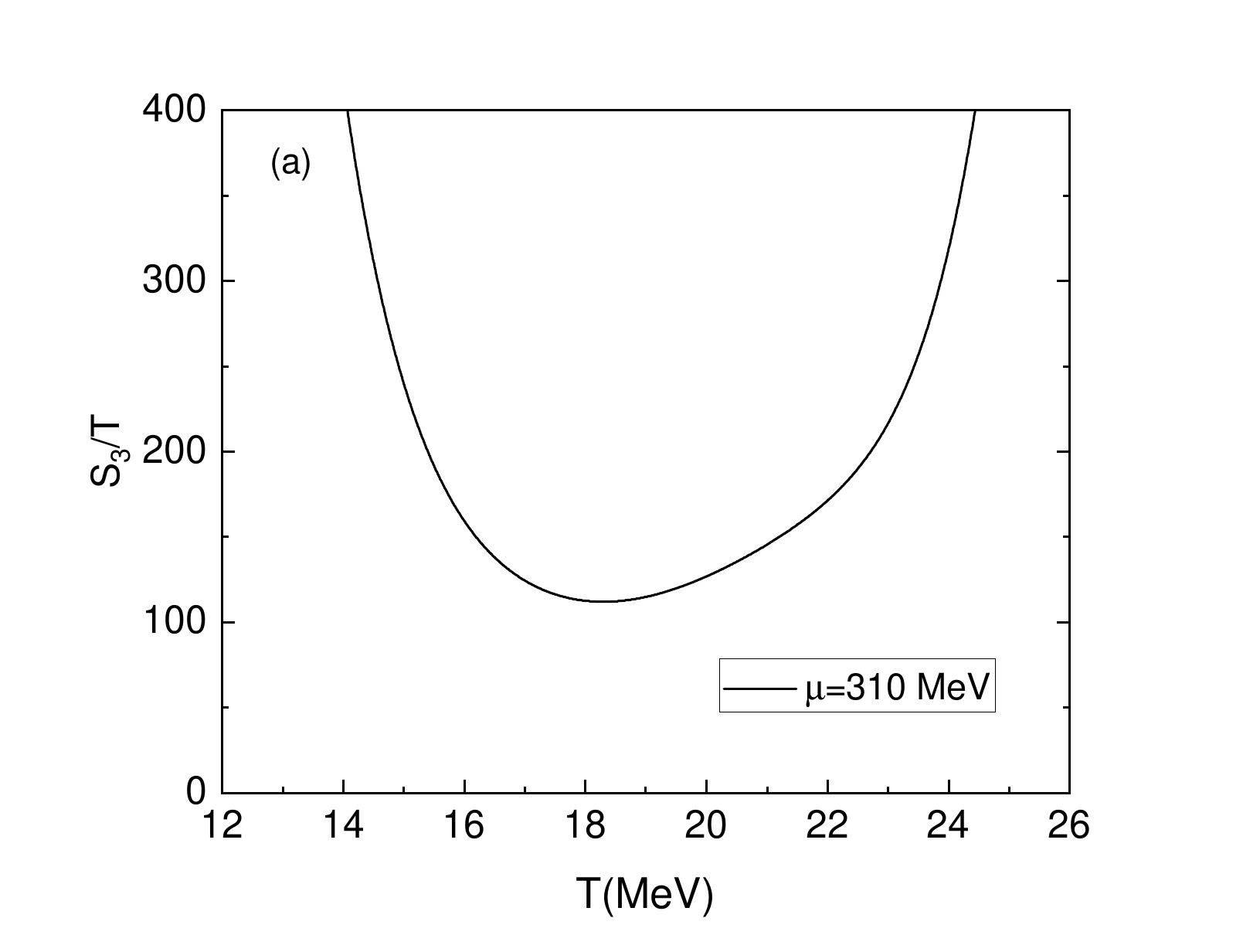}\hspace*{0.01cm} \epsfxsize=9.0 cm
\epsfysize=6.5cm \epsfbox{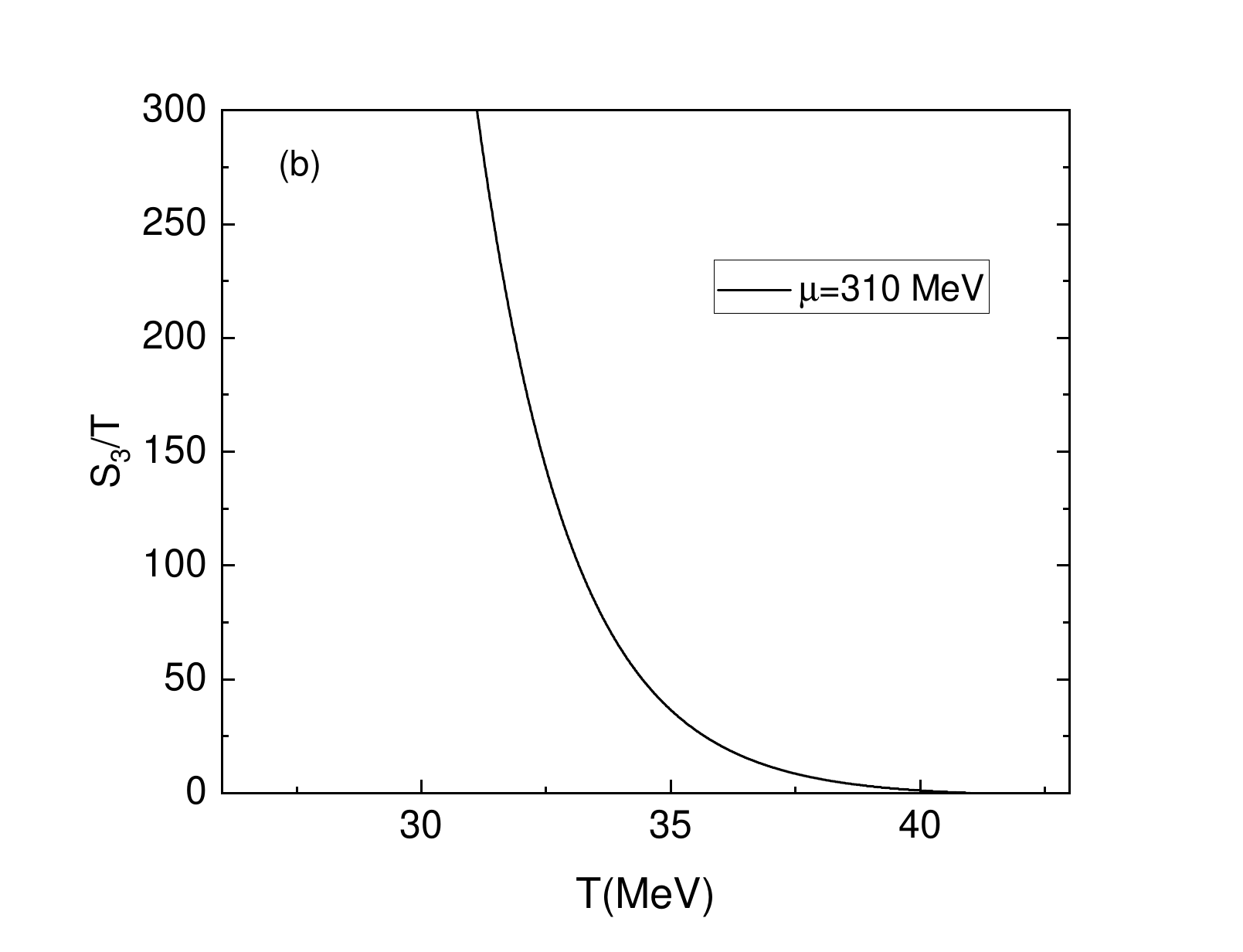}
 \caption{(Color online) (a) The saddle-point action evaluated on the bounce as a function of temperature $T$ when $T< T_{\chi}^c$ for $\mu=310$ MeV. (b) The saddle-point action evaluated on the bounce as a function of temperature $T$ when $T> T_{\chi}^c$ for $\mu=310$ MeV.}
\label{Fig12}
\end{figure}

When fixing the chemical potential at $\mu=310$ MeV, the resulting plot of the $S_3/T$ action as a function of the temperature $T$ due to the appearance of the bounce are shown in Fig.\ref{Fig12} for both cases of the hadron phase and quark phase. From the left panel of Fig.\ref{Fig12}, for a strong first-order phase transition, the $S_3/T$ action reduces firstly to a minimal point at $T\sim 18$ MeV, then it will rise up quickly and intend to diverge again as $T\rightarrow T_{\chi}^c$. The non-monotonic behavior of the $S_3/T$ action is one of the special character of a strong first-order phase transition, which has been also reported in recent studies on a strong cosmological first-order phase transition \cite{Wang:2020jrd} and a strong hadron quark phase transition \cite{Wang:2023omt,Zhou:2020bzk}. Moreover, since the $S_3/T$ action evaluated on the bounce persistently satisfies the requirement of $S_3/T\gg 1$ in the hadron phase, the nucleation rate of the hadron state inside the homogeneous quark phase has been exponentially suppressed in the whole region of the conversional hadron phase, even though the temperature is below the critical coexistence line at $T=T_{\chi}^c$. This is a meaningful result because if the system is cooling down from a very high energy, where the initial state is well prepared as a free quark, the system is more like to remain in a quark state rather than a hadron state in spite of the fact that the temperature is below the critical coexistence line at $T=T_{\chi}^c$ for $\mu>\mu_c$. In other words, for an area of a strong first-order phase transition in phase diagram, it is better to treat the system as a quark matter instead of a hadron matter when $T\leq T_{\chi}^c$.       

When the temperature is above the critical coexistence line at $T=T_{\chi}^c$, the hadron quark phase conversion has been changed to a weak first-order phase transition by reason of the disappearance of the potential barrier as $T\rightarrow T_{\mathrm{sp}}$. In the right panel of Fig.\ref{Fig12}, the $S_3/T$ action as a function of the temperature $T$ due to the appearance of the bounce is illustrated when the chemical potential is fixed at $\mu=310$ MeV. Now it is a weak first-order phase conversion, the evolution of the $S_3/T$ action with the temperature exhibits a similar result as the prior cases at $\mu=306$ MeV. The $S_3/T$ action tends to become divergent when the temperature approaches to the critical one at $T=T_{\chi}^c$, however, in the opposite direction, as the temperature inclines to the spinodal line, it will descent to the zero dramatically. Unless the temperature is very close to the spinodal line, the action fulfils the a critical condition such that $S_3/T>1$. As a result the hadron quark phase transition dose not take place exactly at the critical temperature but up to a certain temperature nearby the spinodal line.           

\section{Summary}

In this work, we have investigated the dynamics of a first-order phase transition via homogeneous thermal nucleation within the Polyakov quark-meson model at finite temperature and density. In the framework of the mean field approximation, the effective potential with the inclusion of a fermionic vacuum term at finite temperature and chemical potential has been calculated and the phase diagram together with the two metastable regions has been obtained. It is found that at low density the chiral phase transition is a crossover, however it will terminate and change into a first-order phase transition at high chemical potential around the critical end point (CEP). By using a geometric method of the effective potential, we can precisely locate the position of the CEP at $(T_{\mathrm{E}},\mu_{\mathrm{E}})\simeq(301.4 \mathrm{MeV}, 62.1 \mathrm{MeV})$. It is worth to pointing out that there is a large uncertainty about the position of CEP for theoretical studies, it is scattered over the region of the baryonic chemical potential at $\mu_B=200\sim 1100$ MeV \cite{Luo:2017faz,Pandav:2022xxx}. Similar to the QM and PQM models, both the NJL and PNJL models predict a very large chemical potential at $\mu=\mu_B/3\simeq 300$ MeV, whereas the functional methods, such as the functional renormalization group (FRG) approach and the Dyson-Schwinger (DS) equations, give out their critical value only around at $\mu_B=200\sim 220$ MeV \cite{Fu:2022gou}. Since Lattice QCD calculations are not so reliable at high density, to identify and locate the CEP in the heavy-ion collisions experiment are very crucial and important. Furthermore, in the region of a first-order phase transition, besides the coexistence line at $T=T_{\chi}^c$, two additional spinodal lines which are usually ignored in the standard phase diagram of QCD have been also presented explicitly. From phase diagram in Fig.\ref{Fig04}, we can find the distance of these two spinodal lines is to decline with the temperature descent. In the end, the spinodal lines will joint together with the coexistence line and terminate by the same point at the CEP, where the order of phase transition is of second order.        

In order to study the instanton tunneling between two vacua, we need to search all minima of the effective potential. Aside from a standard way through the calculation of the Hessian matrix, we have provided an alternative method by constrained the effective potential to the mesonic field direction and the Polyakov-loop field direction in the parameter space for the sake of simplification. The advantage of the our geometric method is that it supplies an intuitive and vivid way to find two minima nearby which can actually have the tunneling solution. Moreover, we can change the complicated N-dimensional problem into a typical one-dimensional problem and this will make us much easy to solve the equation of motion to get the bounce solution both in an exactly numerical method and an analytically thin-wall approximation. Of course, the method in the present study have two main shortages. For the first one, the procedure of finding the minima is usually trouble and sophisticated, especially when the potential has many order parameter variables and many minima, since we have to explore all minima of the effective potential one by one. For the second one, there is an opportunity to miss some minima around a specific point of the effective potential, but the absent minima can be recovered again if we run over all the established minima. Fortunately, although there are three order parameters in the PQM model, the effective potential of the model has merely two minima and the instanton tunneling is more likely to happen in the $\sigma$ field direction rather than the Polyakov-loop field directions. Accordingly, the system of the three coupled equation of motions for the bounce has been dissociated, and what should be solved is an ordinary differential equation of the $\sigma$ field in Eq.(\ref{eom_s}). Thus the problem has been simplified completely and the method developed in a single field can be applied directly.        

With the appropriate boundary conditions, the exact bubble profiles were numerically calculated for a first-order phase transition. For convenience, our discussions have been separated into a weak first-order hadron quark conversion and a strong one. For both cases, when the temperature is close to the critical coexistence line at $T=T_{\chi}^c$, the bubble profile exhibits a core structure. The kernel of the bubble stays in a true vacuum at $\sigma=\sigma_T$, while it is separated from a homogeneous false vacuum at $\sigma=\sigma_F$ by a thin wall. On the other side, when the temperature leaves far away from the critical one, the bubble profile tends to display a coreless structure since the thickness of the bubble wall has the same order of its radius. To give a more detailed information about the size of the bounce, the typical radius has been roughly estimated by the maximal value of the first derivative of the bubble profile. For a weak first-order phase transition, as $T\rightarrow T_{\chi}^c$, the radius $R_c$ goes to the infinite, whereas it will drop down to zero very quickly as the temperature approaches to the spinodal line. For a strong first-order phase transition as $\mu>\mu_c$ and $T<T_{\chi}^c$, the radius becomes divergent as $T\rightarrow T_{\chi}^c$, but it remains almost a constant with the descent of the temperature. This is one of special features of a strong first-order phase transition.           

The surface tension plays an important role during the procedure of bubble nucleation because it is an amount of the energy cost per unit area in production of the interface between two vacua. In the scenario of a weak first-order phase transition, the surface tension reaches to a relative larger value at $T=T_{\chi}^c$, then it will continuously reduce to zero as $T\rightarrow T_{\mathrm{sp}}$. on the other hand, in the context of a strong first-order phase transition, the surface tension is likely to keep its constant value as $T\rightarrow 0$ since there is no spinodal temperature any longer. Just like the QM model \cite{Wang:2023omt}, the inclusion of the deconfinement effect does not change the surface tension dramatically, it remains at a very small value below $4$ MeV. This result is consistent with the predicts of most QCD effective models, such as the MIT bag model \cite{Oertel:2008wr}, the Friedberg-Lee model\cite{Zhou:2020bzk}, the chiral nucleon-mesonw model\cite{Fraga:2018cvr} and NJL model \cite{Ke:2013wga,Garcia:2013eaa,Xia:2022tvx}. Such a mall value of the surface tension would not only favor a mixed phase in the cores of neutron stars but also provide a possibly observable signal of the QCD first-order phase transition during the core-collapse supernova explosions in astrophysics.

Similarly, the $S_3/T$ action evaluated on the bounce solution shows a common characteristic for a weak first-order phase transition. It is divergent as the critical temperature closes in, while in the opposite direction, it will fall down dramatically to zero as $T\rightarrow T_{\mathrm{sp}}$. Considered the exponential dependence of the nucleation rate on the $S_3/T$ action, the decay of the false vacuum is still exponentially suppressed when $S_3/T>1$. Hence the false vacuum could live as a metastable state for a relatively long time so long as the system lies between the up and low spinodal lines. Only when the temperature is very close to the spinodal critical line, the $S_3/T$ will go through the unity $1$. As a result, the phase boundary for a weak first-order phase transition should be resized accordingly with the results in the present work. Moreover, for a strong first-order phase transition, the situation will become more worse, since there is no down spinodal line, the nucleation rate of the true vacuum is always exponentially suppressed as $S_3/T\gg 1$. In other words, a “conventional” hadron matter below the critical coexistence line for $\mu>\mu_c$ should be potentially treated as a quark matter either, such a conclusion together with the low values of the surface tension will favor a more complicated structure of strong interaction matter in high density.

\begin{acknowledgments}
We thank Ken D. Olum and Ziwang Yu for many fruitful comments and discussions. This work is supported in part by National Natural Science Foundation of China (NSFC) under No.11675048.
\end{acknowledgments}

\end{document}